\documentclass[conference]{IEEEtran}
\IEEEoverridecommandlockouts
\usepackage{cite}
\usepackage{amsmath,amssymb,amsfonts}
\usepackage{algorithmic}
\usepackage{graphicx}
\usepackage{textcomp}
\usepackage{xcolor}

\usepackage{url}
\usepackage{enumitem}
\usepackage{multirow}
\usepackage{subcaption}
\usepackage{booktabs}
\usepackage[caption=false, font=footnotesize]{subfig}
\usepackage{threeparttable}
\usepackage{longtable}
\usepackage[most]{tcolorbox}
\usepackage{todonotes}

\newenvironment{mybox}{
    \begin{tcolorbox}[boxsep=2pt,left=0pt,right=0pt,top=0pt,bottom=0pt]}{ 
    \end{tcolorbox}}
    
\def\BibTeX{{\rm B\kern-.05em{\sc i\kern-.025em b}\kern-.08em
    T\kern-.1667em\lower.7ex\hbox{E}\kern-.125emX}}
\begin{document}

\title{Understanding Web Application Workloads and Their Applications: Systematic Literature Review and Characterization}

\author{
    \IEEEauthorblockN{Roozbeh Aghili, Qiaolin Qin, Heng Li, Foutse Khomh}
    \IEEEauthorblockA{Polytechnique Montreal, Canada
    \\\{roozbeh.aghili, qiaolin.qin, heng.li, foutse.khomh\}@polymtl.ca}
}

\maketitle

\begin{abstract}
Web applications, accessible via web browsers over the Internet, facilitate complex functionalities without local software installation. In the context of web applications, a workload refers to the number of user requests sent by users or applications to the underlying system. Existing studies have leveraged web application workloads to achieve various objectives, such as workload prediction and auto-scaling. However, these studies are conducted in an \textit{ad hoc} manner, lacking a systematic understanding of the characteristics of web application workloads. In this study, we first conduct a systematic literature review to identify and analyze existing studies leveraging web application workloads. Our analysis sheds light on their workload utilization, analysis techniques, and high-level objectives. We further systematically analyze the characteristics of the web application workloads identified in the literature review. Our analysis centers on characterizing these workloads at two distinct temporal granularities: daily and weekly. We successfully identify and categorize three daily and three weekly patterns within the workloads. By providing a statistical characterization of these workload patterns, our study highlights the uniqueness of each pattern, paving the way for the development of realistic workload generation and resource provisioning techniques that can benefit a range of applications and research areas.
\end{abstract}


\section{Introduction}
\textit{Web applications} are delivered via the World Wide Web to users, allowing them to access complex functionality without installing or configuring local software components (except a browser). In the context of web applications, the term \textit{workload} refers to requests sent by users or applications to the underlying system. Web applications, such as Wikipedia~\cite{wikipedia}, typically monitor the workload data for various purposes, such as user behavior analysis (e.g.,~\cite{urdaneta2009wikipedia, chowdhury2014category}) or resource allocation (e.g.,~\cite{calheiros2011virtual, zhou2022cushion}). Studying and analyzing these workloads are crucial in understanding the dynamics of user interactions, server responses, and resource utilization for web applications.

Web application workloads are valuable not only for workload prediction (e.g.,~\cite{amekraz2018adaptive, roy2011efficient}), auto-scaling (e.g.,~\cite{dogani2022k, imdoukh2020machine}) and the development of self-healing systems (e.g.,~\cite{moreno2018flexible, moreno2016efficient}), but also for software maintenance activities such as performance optimization (e.g.,~\cite{bairavasundaram2004x, barford1999changes}) and capacity planning (e.g.,~\cite{shi2023auto, abdullah2020predictive}). Previous efforts have also included literature reviews and survey studies that summarize advancements in closely related areas of workload characterization (e.g.,~\cite{calzarossa2016workload, shishira2017workload}), auto-scaling (e.g.,~\cite{singh2019research, qu2018auto}), and workload prediction (e.g.,~\cite{masdari2020survey}). 

Despite the significance of workload data in employing and evaluating these techniques, no existing work has systematically studied the characteristics of web application workloads and their applications. Therefore, in this study, we undertake a \textit{Systematic Literature Review (SLR)} to identify existing studies utilizing web application workloads.  From our SLR, we identify 12 web application workload datasets (worth more than 8.5 years of workloads in total) and study their applications. We then systematically characterize these workloads, providing valuable insights for future research on software maintenance practices, such as realistic workload generation and resource provisioning strategies. Our two \textit{Research Questions (RQ)} are as follows.

\begin{itemize}

\item[RQ1] \textbf{\textit{How are web application workloads used in existing research?}}
While web application workloads are known to be valuable for various purposes, there is a gap in understanding their usage. This RQ aims to bridge this gap by identifying diverse applications of these workloads. To achieve this, we conduct an SLR examining articles that utilize web application workloads. This SLR involves a comprehensive search across two research databases, followed by article selection and forward and backward snowballing, ultimately resulting in a dataset of 78 articles. We analyze how these workloads are used, revealing current trends and potential limitations.

\item[RQ2] \textbf{\textit{What are the existing patterns in web application workloads?}}
Although web application workloads are crucial for tasks such as workload prediction and auto-scaling, their inherent characteristics remain largely unexplored. This RQ focuses on systematically characterizing these workloads and identifying recurring patterns. Through a thorough examination of web application workloads using clustering techniques, we offer insights that researchers and practitioners can use to develop more accurate and efficient tools and techniques that are customized to the specific characteristics of workload patterns.
\end{itemize}

\begin{figure}[t!]
  \centering
  \includegraphics[width=.48\textwidth]{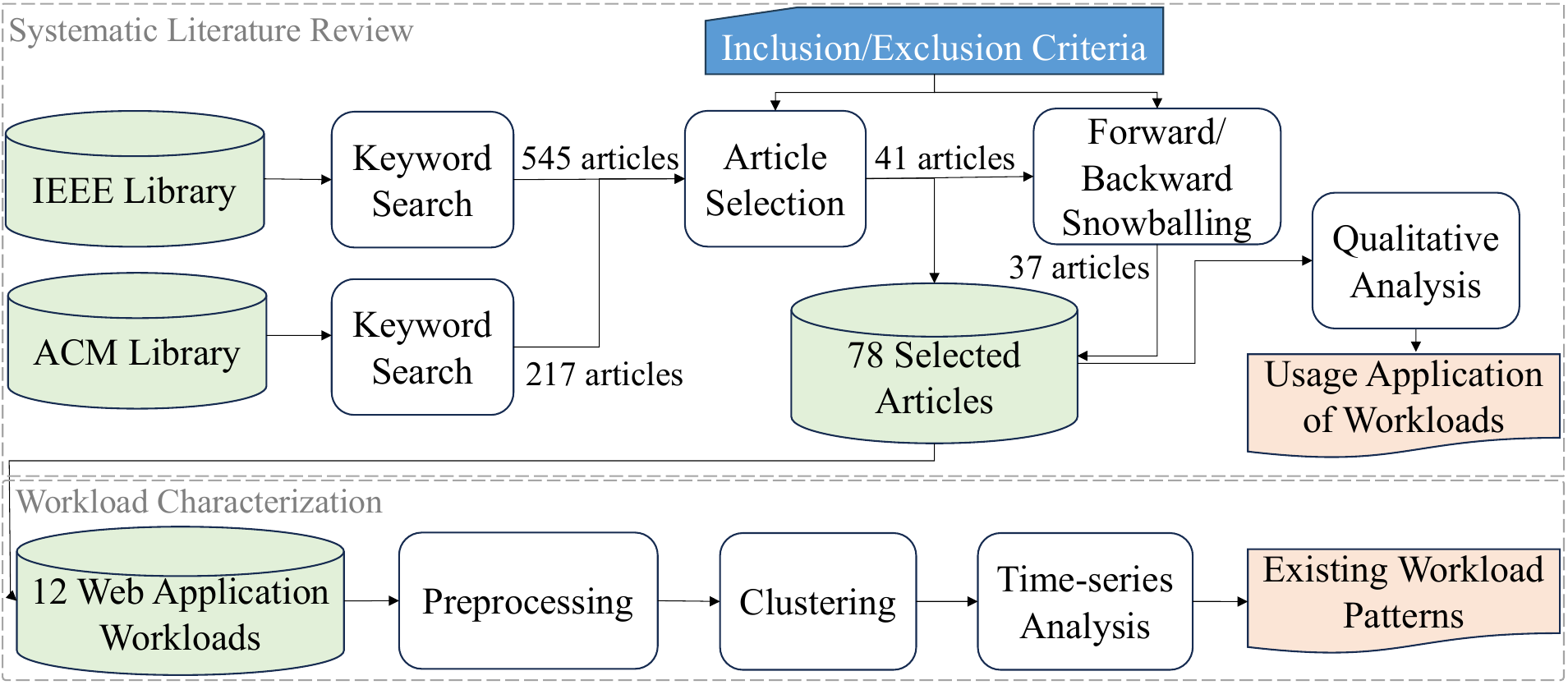}  \caption[Short text]{{Overview of our study}}
  \label{fig:overview}
\end{figure}

Figure~\ref{fig:overview} provides an overview of our study. Our research is initiated by the identification of 78 articles published between 1995 and 2024. These research articles were selected because they utilized web application workloads in their studies. To identify these papers, we perform a systematic literature search involving three stages: a thorough exploration of research databases, article selection, and snowballing the selected articles to find relevant papers. Once the set of articles is finalized, we identify the web applications they have utilized. We extract the available workloads and employ clustering techniques to characterize these workloads. Characterizing workload patterns of web applications is of importance, offering several advantages, including (1) uncovering hidden patterns among web application workloads, (2) revealing specific characteristics associated with each workload pattern, and (3) enabling the development of load testing tools, workload generators, and resource provisioning techniques based on our findings. 

Our work makes several important contributions:

\begin{enumerate}
    \item 
    Our SLR offers a comprehensive insight into the applications of web application workloads. 
    \item We identified a total of 12 publicly available web application workload datasets through our SLR.
    \item We identified three daily and three weekly workload patterns across different web applications and presented their distinct characteristics. The results of this analysis could guide future work on realistic workload generation and resource provisioning.
\end{enumerate}

We share our replication package\footnote{\url{https://github.com/mooselab/web-app-workloads}} so that future work can replicate or extend our study. 


\section{Related Work}
\label{sec:related_work}

This work performs 
an SLR on the usage of web application workloads and explore the characteristics of these workloads. 
Thus, we discuss related work on the following two aspects.

\subsection{Systematic Literature Reviews in Similar Context} While no previous work has systematically investigated the usage of web application workloads in the literature, there are existing SLR studies and survey papers on similar domains, including workload characterization (e.g.,~\cite{calzarossa2016workload, shishira2017workload}), auto-scaling (e.g.,~\cite{singh2019research, qu2018auto}), and workload prediction (e.g.,~\cite{masdari2020survey}). Several articles within these domains have leveraged web application workloads to implement and evaluate their systems. 

Calzarossa et al.~\cite{calzarossa2016workload} presents a survey of the state of the art of workload characterization. They divide workloads in five areas: web applications, social networks, video services, mobile devices, and cloud computing. In the section related to web application workloads, they study 31 research articles. They further divide the web workloads into three sub-domains of shopping services, web robot traffic, and web content. They conclude that workload characterization is a fundamental component for all these three domains. Similarly, Shishira et al.~\cite{shishira2017workload} categorize the workloads into four areas of web applications, social networks, video services, etc.

Auto-scaling is the dynamic adjustment of the number of resources allocated to an application based on its workload, ensuring optimal performance and cost efficiency~\cite{qu2018auto}. In~\cite{qu2018auto}, Qu et al. review the advancements of web application auto-scaling in cloud systems. While providing a taxonomy and survey, this study does not focus on the specific input data the articles utilized for their auto-scaling techniques. A more recent survey by Singh et al.~\cite{singh2019research} in the auto-scaling domain reports the experiment phase and the data used by the studied articles. They observe that some articles in their study use web application workloads, while others rely on cloud workloads, synthetic workloads, simulations, or custom testbeds.

Masdari et al.~\cite{masdari2020survey} provide a survey of the workload prediction methods. Based on their categorization, the studied papers either use realistic web application workloads, cloud workloads, self-collected workloads, or simulation techniques. Their findings reveal that over 86\% of the articles exclusively utilize a single workload dataset to assess their results.

\subsection{Workload Characterization} 
Previous research efforts have characterized the cloud-based batch-job workloads such as the Google cluster trace (e.g., \cite{shekhawat2018datacenter, patel2020clustering}) and the Alibaba cluster trace (e.g., ~\cite{luo2021characterizing, chen2018does}). One of the common approaches to characterize workloads is utilizing clustering techniques. Patel et al.~\cite{patel2020clustering} apply clustering approaches to the Google cluster trace~\cite{google} and Bitbrains workloads~\cite{bitbrains}, comparing K-Means and Gaussian Mixture Model performance for cluster representativeness. Similarly, Shekhawat et al.~\cite{shekhawat2018datacenter} employ K-Means and six other machine learning algorithms to characterize the workload patterns in the Google cluster trace~\cite{google} and Bitbrains workloads~\cite{bitbrains}, considering CPU usage, memory usage, disk usage, and network usage as their workload parameters. In another study, Chen et al.~\cite{chen2018does} cluster workloads in the Alibaba cluster trace~\cite{alibaba} using K-Means and workload features such as job duration, CPU cores, memory utilization, and disk utilization, providing a statistical profile of workload behavior. 

In the context of web application workloads, Shahidinejad et al.~\cite{shahidinejad2021resource} combine K-Means and the \textit{Imperialist Competitive Algorithm (ICA)} to classify and analyze web application workloads, including Worldcup98~\cite{Worldcup98} and NASA~\cite{nasa}. In another work, Chowdhury et al.~\cite{chowdhury2014category} employ time-series analysis techniques to characterize the YouTube workload~\cite{youtube}. 

While existing studies have characterized web application workloads (e.g., \cite{shahidinejad2021resource, arlitt1996web, chowdhury2014category}), they often lack a comprehensive overview of the general characteristics of different workloads. Unlike prior research, focusing on one or two specific workloads, our approach involves identifying and analyzing 12 public web application workloads resulting from an SLR. 
With this approach, we aim to uncover broad and consistent patterns across these diverse workloads, offering a more comprehensive understanding of web application workload dynamics.

\section{Systematic Literature Review}
\label{sec:slr}

\subsection{Dimensions of the Review}
\label{sec:dimensions}
The goal of this review is to provide researchers and practitioners with a structured overview of existing research using web application workloads. Through a meticulous examination of the research literature, we derive several key dimensions. These dimensions provide objective descriptions of different techniques and objectives and organize the literature review. The identified dimensions are as follows.

\begin{itemize}
    \item \textit{Workloads:} Which web application workloads have been utilized in the literature?
    \item \textit{Techniques and Objectives:} What are the overall research objectives of the literature and what types of techniques have been used to achieve them?
    \item \textit{Trends:} What are the temporal trends in workload usage and research objectives over time?

\end{itemize}

\subsection{Approach}
\label{sec:slr_approach}

Through a comprehensive search across two research databases, we identify 762 potential candidate articles.  After applying inclusion and exclusion criteria, we narrow down the selection to 41 papers. To ensure a comprehensive review, we employ forward and backward snowballing techniques, adding 33 more articles to our selection. The article selection methodology aligns with the systematic approach recommended by Wohlin et al.~\cite{wohlin2012experimentation} and is described in the following.

\subsubsection{Searching Research Databases} 
\label{sec:search_databases}
We conduct our literature search on two highly reputable research databases: IEEE library~\footnote{\url{https://ieeexplore.ieee.org/}} and ACM library~\footnote{\url{https://dl.acm.org/}}. We aim to identify papers from both conference proceedings and journals, written in English and published between January 2014 and December 2023, a time span of 10 years. To pinpoint the articles that either introduce or utilize web application workloads, we use the following query to search the databases:

\textit{[[Title: web application] OR [Title: workload] OR [Title: trace]] AND [Abstract: dataset] AND [E-Publication Date: (2014/01/01 TO 2023/12/31)]}

Upon conducting this query on the IEEE library, we identify a total of 545 articles that satisfy our criteria. Simultaneously, our search from the ACM library yields an additional 217 articles after deduplication. Combining papers from both sources, we successfully compile a total of 762 candidate articles.

\subsubsection{Article Selection} 
\label{sec:article_selection}
We screen each paper, beginning with its abstract, and then proceed through the entire article until a decision is made based on the inclusion and exclusion criteria.

\underline{Inclusion criteria:}
\begin{itemize}
    \item Papers must be written in English.
    \item Papers must be in conference proceedings or journals.
    \item Papers must introduce or utilize web application workloads that are publicly available and representative of real-world applications.
    \item The utilized workloads must include timestamp information, as this data will enable us to conduct subsequent analysis and characterization of the workloads.
\end{itemize}

\underline{Exclusion criteria:}
\begin{itemize}
    \item Papers not publicly available.
    \item Papers introducing or utilizing private data (e.g., with \textit{Non-Disclosure Agreement (NDA)} contracts).
    \item Papers introducing or utilizing non-web application data such as cloud-based batch-jobs (e.g., Google cluster trace), social networks data, mobile devices workloads, and synthetic workloads. Overall, we exclude articles that leverage data unrelated to user interaction workloads.
\end{itemize}

After performing the article selection step, we end up selecting 41 out of the 762 candidate articles.

\begin{table*}[t]
\caption{List of web application workloads of different literature}
\begin{threeparttable}
    \small
    \resizebox{1\textwidth}{!}{
    \begin{tabular} {p{0.02\textwidth} p{0.09\textwidth} p{0.07\textwidth} p{0.09\textwidth} p{0.53\textwidth} p{0.04\textwidth} p{0.15\textwidth}}
    \toprule
    \textbf{ID} & \textbf{Workload} & \textbf{Duration~\textsuperscript{a}} & \textbf{\# Instances} & \textbf{Description} & \textbf{Freq.} & \textbf{Ref. examples~\textsuperscript{d}} \\
    \toprule
        1 & Wikipedia~\textsuperscript{b} & 5.5 years & 1.0 T & Wikipedia workload consists of the number of users accessing Wikimedia Foundation articles from January 2018 to August 2023. & 23 & \cite{amekraz2018adaptive, wu2022long, jayakumar2022cloudbruno} \\
        \midrule

        2 & Calgary & 1 year & 727 K & Calgary workload includes one year of HTTP requests to the University of Calgary's Department of Computer Science server in 1994-1995. & 13 & \cite{kumar2021performance, kumar2020cloud, ebadifard2021autonomic} \\
        \midrule

        3 & Saskatchewan & 7 months & 2.4 M & This workload contains seven months of HTTP requests sent to the University of Saskatchewan's server in Canada in 1995. & 18 & \cite{kumar2020cloud, saravanan2024workload, karthikeyan2023cosco2} \\
        \midrule
        
        4 & Boston & 6 months & 1.1 M & Boston workload contains HTTP requests sent to the Boston University Computer Science Department from November 1994 to May 1995. & 4 & \cite{cunha1995characteristics, crovella1997self} \\
        \midrule

        5 & Retailrocket & 4.5 months & 2.8 M & This workload contains HTTP requests to servers of an anonymous real-world e-commerce website over a period of 4.5 months in 2015. & 1 & \cite{bauer2018chameleon} \\
        \midrule
        
        6 & Worldcup98 & 2 months & 1.3 B & Worldcup98 workload consists of all requests made to the 1998 World Cup website between 30 April 1998 and 26 July 1998. & 29 & \cite{dogani2022k, bauer2018chameleon, tahir2020online} \\
        \midrule
        
        7 & NASA & 2 months & 3.5 M & NASA workload contains the requests sent to NASA Kennedy Space Center in Florida, USA, between 1 July 1995 and 31 August 1995. & 35 & \cite{dogani2022k, shahidinejad2021resource, amekraz2022canfis} \\ 
        \midrule

        8 & YouTube & 44 days & 1.5 M & A collection of workloads from a campus network measurement on YouTube traffic between June 2007 and March 2008 spans 10 months, though the actual data covers a period of 44 days. & 3 & \cite{chowdhury2014category, zink2009characteristics, zink2008watch} \\
        \midrule

        9 & Madrid & 1 month & \_\_~\textsuperscript{c} & Real web service logs from the Complutense University of Madrid. This dataset was collected hourly throughout the month of May 2018. & 1 & \cite{moreno2019efficient} \\
        \midrule
        
        10 & ClarkNet & 14 days & 3.3 M & Clarknet was an Internet service provider located in Maryland, USA. This workload consists of two weeks of data in September 1995. & 18 & \cite{tahir2020online, abdullah2020burst, zhou2022cushion} \\
        \midrule
        
        11 & EPA & 1 day & 47.7 K & EPA workload contains one day of requests sent to the EPA server located at North Carolina, USA in August 1995. & 1 & \cite{prevost2011prediction} \\
        \midrule
        
        12 & SDSC & 1 day & 28.3 K & SDSC workload includes requests to the servers of the San Diego Supercomputer Center in California over a single day in August 1995. & 1 & \cite{feitelson2006metrics} \\

        \bottomrule
 \end{tabular}
    }
        \begin{tablenotes}
        \footnotesize
        \item[a] This indicates the duration of each workload dataset. Certain literature utilized segments of these durations.
        \item[b] Unlike other workloads that have one specific dataset, Wikipedia offers a free API allowing users to download customized data sets. The duration and \\ instances mentioned here are the data extracted from Wikipedia for this study.
        \item[c] Unlike other workloads, this dataset does not provide raw data; instead, it offers preprocessed data.
        \item[d] The full list of articles is available in our replication package.
        \end{tablenotes}
    \end{threeparttable}
    \label{tab:workloads}
\end{table*}

\begin{table*}[t]
\caption{List of different literature objectives with their corresponding techniques}
    \centering
    \small
    \resizebox{1\textwidth}{!}{
    \begin{tabular}{p{0.22\textwidth} p{0.55\textwidth} p{0.04\textwidth} p{0.15\textwidth}}
    \toprule
    \textbf{Objective} & \textbf{Utilized Techniques} & \textbf{Freq.} & \textbf{Ref. examples} \\
    \toprule
    \textbf{Resource Management} & \textbf{} & \\

    \quad Load Balancing & Classical Machine Learning, Deep Learning, Time-series/Statistical Analysis, Queueing Theory & 4 & \cite{lei2003research, ebadifard2021autonomic} \\

    \quad Resource Provisioning & Classical Machine Learning, Deep Learning, Optimization, Control Theory, Fuzzy Logic, Queueing Theory & 6 & \cite{shahidinejad2021resource, zhou2022cushion} \\

    \quad Caching Optimization & Time-series/Statistical Analysis & 2 & \cite{bairavasundaram2004x, barford1999changes} \\
    \midrule
    
    \textbf{Workload Analysis} & \textbf{} & \\

    \quad Workload Prediction & Time-series Models, Classical Machine Learning, Deep Learning, Optimization, Time-series/Statistical Analysis, Filtering and Signal Processing, Markov Models, Fuzzy Logic & 44 & \cite{bauer2018chameleon, kim2018cloudinsight, karthikeyan2023cosco2} \\

    \quad Workload Classification & Time-series Models, Classical Machine Learning, Optimization, Time-series/Statistical Analysis & 4 & \cite{herbst2013self, shahidinejad2021resource, urdaneta2009wikipedia} \\

    \quad Workload Characterization & Optimization, Time-series/Statistical Analysis & 13 & \cite{jaiman2018heron, arlitt2000workload, arlitt1996web} \\
    \midrule
    
    \textbf{Self Adaptation} & \textbf{} & \\
    
    \quad Auto-scaling & Optimization, Control Theory, Fuzzy logic, Queueing Theory & 14 & \cite{bauer2018chameleon, dogani2022k} \\

    \quad Self-healing & Optimization, Markov Model & 2 & \cite{moreno2018flexible, moreno2016efficient} \\
    \midrule
    
    \textbf{Benchmarking} & Time-series Models, Optimization, Control Theory, Time-series/Statistical Analysis, Queueing Theory & 6 & \cite{kumar2021performancee, kumar2021performance} \\
    \bottomrule
 \end{tabular}
 
    }
    \label{tab:objectives&techniques}
\end{table*}

\subsubsection{Forward and Backward Snowballing} 
\label{sec:snowball}

To ensure the comprehensiveness of our study and to capture potentially relevant articles, we initiate the forward and backward snowballing phase on the selected 41 papers. During this phase, for each of the selected articles, we inspect the list of papers that have cited them (i.e., forward snowballing) and their reference papers (i.e., backward snowballing) and perform the article selection step (as described in Section~\ref{sec:article_selection}) to decide whether the new article matches with our inclusion and exclusion criteria. After performing this step, we end up having 33 new articles. Combining the articles retrieved through our initial search of research databases and the subsequent snowballing process, we obtain a total of 78 articles, each of which introduces or utilizes public web application workloads.

\subsubsection{Qualitative Analysis} 
\label{sec:analysis}
We manually examine each of the selected articles to extract its dimensions (i.e., workloads, techniques, objectives). This labeling process involves reviewing all sections of each paper to identify the relevant information, though most of this information is typically found in the methodology sections. We use an open coding approach~\cite{khandkar2009open} to extract the desired information. To label the articles, the first two authors of the paper (i.e., coders) jointly perform a coding process, determining each article’s workloads, techniques, and objectives. We perform a five-step coding process as follows.

\textbf{Step 1: Coding.} Each coder independently analyzes the first batch of articles (i.e., the initial 37) and assigns labels for each dimension (i.e., workloads, techniques, objectives) of each paper. Multiple labels can be assigned to each dimension.

\textbf{Step 2: Discussion.} The coders share their responses and discuss the labels they created, aiming for a common understanding. We join related labels and refine high-level ones, then revise the coding of the first batch accordingly.

\textbf{Step 3: Coding.} Each coder analyzes the second batch of articles (i.e., the final 37) based on the discussion results.

\textbf{Step 4: Resolving disagreements.} The coders compare their final results from step 3 and discuss any remaining conflicts, attempting to resolve them. If an agreement cannot be reached, the third author makes the final decision.

\textbf{Step 5: Final revision.} In the final stage, we create a mind map from all the produced labels. We then discuss the labels and form a hierarchy, change some labels’ names for clarity, and merge some small categories to be cohesive.


\subsubsection{Measuring the Reliability}
\label{sec:reliability}
Ensuring reliability is crucial for validating coding results \cite{artstein2008inter}. The coding results are reliable when there is a specific level of agreement between coders, referred to as inter-coder agreement. In this study, we employ Cohen’s kappa \cite{cohen1960coefficient} as a metric to quantify the reliability of agreements between two coders. We evaluate our coding reliability for the second batch of the articles (i.e., after the discussion session) and achieve a Cohen's kappa of 0.94. A value of kappa $\geq$ 0.80 indicates a strong agreement~\cite{mchugh2012interrater}.

\subsection{Results}
We present the findings of our SLR in three main parts: the analysis of web application workloads, the discussion of objectives and techniques, and the temporal trend analysis.

\subsubsection{Web Application Workloads}
\label{sec:web_app_workloads}
Table~\ref{tab:workloads} presents workload datasets used in the literature, showcasing details such as duration, data instances, description, frequency of usage in articles, and example references. Our review identifies 12 distinct workloads commonly used in the literature: Wikipedia~\cite{wikipedia}, Worldcup98~\cite{Worldcup98}, NASA~\cite{nasa}, Saskatchewan~\cite{saskatchewan}, Calgary~\cite{calgary}, EPA~\cite{epa}, Clarknet~\cite{clarknet}, Retailrocket~\cite{retailrocket}, Boston~\cite{boston}, SDSC~\cite{sdsc}, Youtube~\cite{youtube}, and Madrid~\cite{madrid}.

Some workloads, such as the NASA dataset, appeared in many articles (35 in total), while others, such as RetailRocket, SDSC, and Madrid, were used much less frequently, each appearing in just one article. These datasets vary significantly in duration and the number of instances. For instance, the WorldCup98 workload, characterized by high demand, consists of 1.3 billion instances. In contrast, the SDSC workload spans only one day with 28 thousand instances.

\begin{mybox}
Existing studies repetitively use 12 public web application workload datasets, covering a wide range of workload duration (from a single day to 5.5 years) and size (from 28.3 thousand to 1.0 trillion instances).
\end{mybox}

\subsubsection{Objectives and Techniques}
Table~\ref{tab:objectives&techniques} presents objectives and primary techniques from literature articles, along with their frequency of usage and example references. A more detailed list of objectives and techniques along with all the reference articles can be found in our replication package\footnote{\url{https://github.com/mooselab/web-app-workloads}}. Below we define the objective categories.

The first objective theme is \textbf{Resource Management}, which aims to ensure the efficient allocation, utilization, and optimization of computing resources within a system or network. It includes three objectives: \textit{Load Balancing}, \textit{Resource Provisioning}, and \textit{Caching Optimization}. 

Load balancing is the distribution of incoming network traffic or computational tasks across multiple resources to optimize resource utilization and ensure high availability. Various approaches are employed for this purpose. For example, Riska et al.~\cite{riska2000analytic} apply queueing theory to evaluate load balancing policies in distributed multi-server systems, modeling each server as a queue using \textit{Markov chains}.

Resource provisioning is the process of initially allocating the resources and services from a cloud provider to a customer to meet the requirements of applications. Shahidinejad et al.~\cite{shahidinejad2021resource} employ classical machine learning models to design their Resource Provisioning system. They propose a system to first cluster the workloads using K-Means and then use~\textit{Decision Tree Regression (i.e., DTR)} to determine scaling decisions for efficient resource provisioning. In another work, Zhou et al.~\cite{zhou2022cushion} combines deep learning models (i.e., \textit{Long Short-Term Memory (i.e., LSTM)}) with classical machine learning models (i.e., XGBoost) to provide proactive resource provisioning. They center their work on microservice systems.

Caching optimization is the techniques and strategies aimed at improving the efficiency and effectiveness of caching mechanisms. Time-series and statistical analysis are the main techniques used for this category. For example, Bairavasundaram et al.~\cite{bairavasundaram2004x} create an image of the file system cache using inferred disk traffic data and propose an array caching mechanism.

The second objective theme is \textbf{Workload Analysis}. It is the examination and understanding of patterns, trends, and characteristics of system or network usage. This objective also includes three objectives: \textit{Workload Prediction}, \textit{Workload Classification}, and \textit{Workload Characterization}.

Workload prediction involves using historical data and trends to forecast future workload demands and patterns, aiming to predict resource requirements and maintain optimal system performance. 
This objective represents the most common theme across our literature articles, comprising 45\% of the total objectives outlined in these papers. Due to the number of articles aiming to improve workload prediction approaches, various techniques have been used. Kim et al.~\cite{kim2018cloudinsight} build an ensemble architecture encompassing various time-series models such as \textit{Weighted Moving Average (i.e., WMA)} and \textit{Auto-Regressive Integrated Moving Average (i.e., ARIMA)}. Their ensemble architecture chooses a subset of these models based on the characteristics of the upcoming workload. Deep learning is another popular solution. Shi et al.~\cite{shi2023auto} use \textit{Deep Reinforcement Learning (i.e., DRL)} and Attia et al.~\cite{attia2019application} use deep learning with a \textit{Differential Evolution (i.e., DE)} algorithm as their optimization technique to predict the workloads. In another work, Saravanan et al.~\cite{saravanan2024workload} use markov models in order to filter redundant historical data before passing them to a \textit{Generative Adversarial Network (i.e., GAN)} model.

Workload classification involves using classification techniques to categorize workloads. In~\cite{ali2013workload}, time-series models such as \textit{Sample Entropy} and classical machine Learning models such as \textit{K-Nearest Neighbors (i.e., KNN)} are employed to classify workloads. In~\cite{urdaneta2009wikipedia}, Urdaneta et al. use time-series analysis techniques to classify Wikipedia workload, emphasizing challenges from its large, heterogeneous dataset.

Workload characterization is the understanding and describing the behavior and characteristics of workloads, including patterns, trends, and variations. The majority of articles within this category rely on time-series and statistical analysis due to the ability of time-series analysis to capture temporal dependencies and identify recurring patterns in workload data.

\textbf{Self Adaptation} is the third objective theme. It is the automatic adjustment of system parameters and resources in response to changing conditions or system faults, aiming to improve performance and reliability without human intervention. It includes two categories of \textit{Auto-scaling} and \textit{Self-healing}.

Auto-scaling is the dynamic adjustment of the number of resources allocated to an application based on its workload, ensuring optimal performance and cost efficiency. While resource provisioning focuses on the initial allocation and management of resources, auto-scaling continuously monitors system metrics and automatically adjusts resource levels to match current demand. Control Theory techniques such as~\textit{Proportional Integrative Derivative (i.e., PID)}~\cite{persico2017fuzzy} have been applied to automatically scale out public-cloud resources, utilizing only the customer's existing knowledge base. In another attempt to create an auto-scaling system, Kumar et al.~\cite{kumar2017qos} use queueing theory and design a system that dynamically performs resource corrections at the virtual machine level by considering both underutilization and over-utilization scenarios. 

Self-healing systems are designed to autonomously detect, diagnose, and recover from failures or performance degradation without human intervention, ensuring continuous operation and reliability. In both works of~\cite{moreno2018flexible} and~\cite{moreno2016efficient}, the authors employ a \textit{Markov model} along an optimization technique (i.e., \textit{Stochastic Programming}) to design a self-healing system compatible with various adaptation objectives.

As the last objective, \textbf{Benchmarking} evaluates the performance, reliability, and scalability of systems by measuring and comparing their performance. In\cite{papadopoulos2016peas}, Papadopoulos et al. propose a performance evaluation framework that utilizes \textit{Control Theory} and optimization to evaluate different auto-scaling systems. In another work, Kumar et al.\cite{kumar2021performancee} evaluate optimization algorithms such as \textit{Genetic Algorithm} and \textit{DE} in the task of workload prediction. In a similar attempt, Kumar et al.~\cite{kumar2021performance} propose a performance evaluation framework that assesses the performance of six prediction models such as \textit{ARIMA} and Exponential Smoothing on Workload Prediction.

\begin{mybox}
Current research covers various objectives across four thematic areas: Resource Management, Workload Analysis, Self-Adaptation, and Benchmarking. These studies employ diverse methodologies, ranging from Classical Machine Learning to Markov Models and Time-series Models.
\end{mybox}

\begin{figure}
\centering
\includegraphics[width=0.48\textwidth]{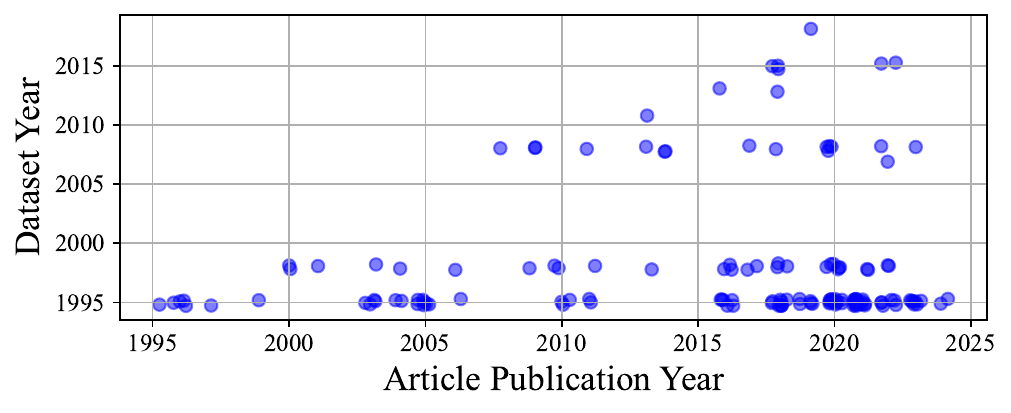}
\caption{Comparison of literature publication years and corresponding workload datasets years}
\label{fig:workload_years}    
\end{figure}

\begin{figure}
\centering
\includegraphics[width=0.48\textwidth]{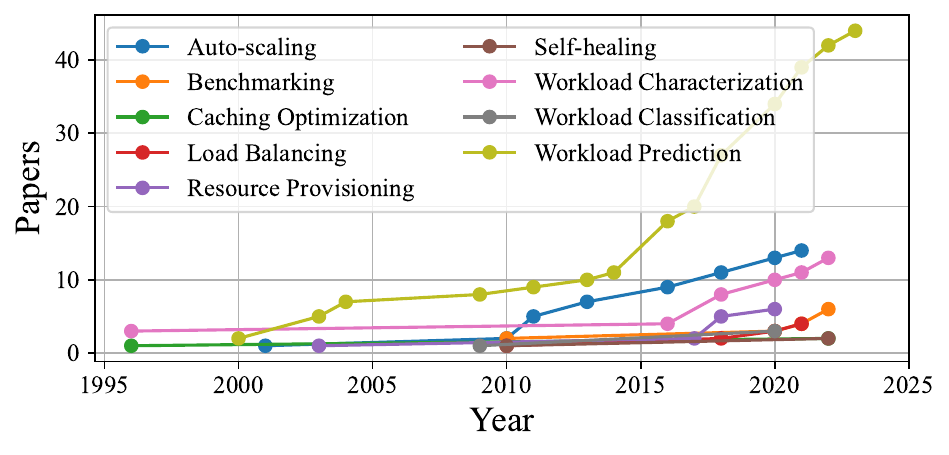}
\caption{Cumulative number of papers per objective over the years}
\label{fig:objective_years}    
\end{figure}

\subsubsection{Temporal Trend Analysis}
Initially, we compare the publication years of the studied articles with those of the workload datasets they employ. Our findings indicate that a significant proportion of the articles utilize relatively dated workload datasets. Specifically, while the median publication year of the articles is 2018, the median publication year of the datasets is 1995. To provide further insight, over 82\% of articles published in 2015 and later rely on datasets that were published in 2000 or earlier. This analysis emphasizes the frequent utilization of older workloads in research studies. Figure~\ref{fig:workload_years} presents the scatter plot of this phenomenon.

Subsequently, we analyze the popularity of different objectives over time. Figure~\ref{fig:objective_years} presents the cumulative number of articles associated with each objective throughout the years. The analysis highlights workload prediction, auto-scaling, and workload characterization as the most prominent objectives. Particularly, there is a noticeable increase in articles focusing on workload prediction, especially after 2014. However, objectives such as self-healing and caching optimization have received relatively less attention, indicating potential areas for further exploration. This observation aligns with the conclusions drawn by Aghili et al.~\cite{aghili2023studying}.

\begin{mybox}
Our trend analysis reveals that Workload prediction has emerged as the primary objective in recent studies. Besides, existing studies rely on significantly dated workload datasets.
\end{mybox}

\section{Workload Characterization}
\label{sec:characterization}

\subsection{Motivation}

Clustering is an essential process for uncovering existing patterns within data. By identifying these patterns in web application workloads, we can gain valuable insights into the behavior and dynamics of web application systems. This understanding is foundational for various tasks, including the development of workload generators and resource provisioning techniques. Leveraging the insights obtained by clustering workload data, researchers and practitioners can design more accurate and efficient tools and techniques that are customized to the specific characteristics of workload patterns. 

\subsection{Approach}
To analyze web application workloads and identify the existing workload patterns, we start by extracting the workloads. We continue by preprocessing through aggregation, standardization, and smoothing. We then analyze the workloads in terms of variability and burstiness. After that, we employ K-Means to cluster the workloads, and finally, we analyze cluster characteristics.

\subsubsection{Web Application Workloads}
As mentioned in Section~\ref{sec:web_app_workloads}, the selected 78 articles utilize 12 web application workloads. A comprehensive description of these workloads, along with key features for each, is presented in Table~\ref{tab:workloads}. 

\subsubsection{Workload Extraction} Most of these 12 web application workloads are accessible for download. However, some, such as Wikipedia, require scripting for data extraction. For the Wikipedia workload, we develop a script to retrieve data spanning 5.5 years, from January 2018 to August 2023.

After gathering all the workloads, we realize that each of them possesses a unique data format. Some are summary-based and only provide information on the number of users who accessed a particular server (e.g., the Wikipedia workload) per time interval, while the others are in the format of log lines and are event-based (e.g., the Worldcup98 workloads). To illustrate this contrast, we provide a comparative example of the raw data from Wikipedia and Worldcup98 workloads in Table~\ref{tab:raw_datasets}. We write separate scripts for each of these data types, extracting the number of users interacting with the web application within one-hour time intervals. 

\subsubsection{Preprocessing the Workloads}
\label{sec:preprocessing}
The preprocessing phase involves three stages: aggregation, standardization, smoothing.

\paragraph{Aggregation} After extracting the user interaction information for all the workloads, we establish two temporal granularities for our analyses: \textit{daily} and \textit{weekly}. These granularities have been widely used for workload-related tasks such as workload generation for load testing and resource provisioning~\cite{fehling2014cloud, jiang2015survey, arlitt2001characterizing}.

\begin{table}
    \caption{Comparing two workload types: summary-based (e.g., Wikipedia) and event-based (e.g., Worldcup98)}
    \small
    \centering
    \begin{tabular}{l}
    \toprule
    \multicolumn{1}{c}{\textbf{Wikipedia - 2023/01/01-00}} \\
    \toprule
    \textit{en.m Cristiano\_Ronaldo 4888 0} \\
    \textit{en.m Lionel\_Messi 2322 0} \\
    \textit{en.m Frédéric\_Chopin 83 0} \\
    \toprule
    \multicolumn{1}{c}{\textbf{Worldcup98}} \\
    \toprule
    \textit{2705258 - - [13/Jul/1998:22:00:01 +0000] "GET/images/} \\
    \textit{102378.gif HTTP/1.0" 200 1658} \\
    
    \textit{1630377 - - [13/Jul/1998:22:00:01 +0000] "GET/images/} \\
    \textit{hm\_score\_up\_line03.gif HTTP/1.0" 200 90} \\
    
    \bottomrule
    \end{tabular}

    \label{tab:raw_datasets}
\end{table}

 \begin{table}
    \caption{The schema of daily and weekly granularities}
    \centering
    \begin{subtable}{0.48\textwidth}
        \centering
        \begin{tabular}{ccccccccc}
        \toprule
             \textbf{Workload} & \textbf{Day} & \textbf{0} & \textbf{1} & \textbf{2} & \textbf{...} & \textbf{22} & \textbf{23} \\
             \toprule
             Wikipedia & 2023-01-01 & 20.7 & 20.6 & 18.5 & ... & 27.7 & 24.9 \\
             Wikipedia & 2023-01-02 & 22.3 & 19.8 & 19.2 & ... & 27.4 & 25.3 \\
             Wikipedia & 2023-01-03 & 21.5 & 19.7 & 19.0 & ... & 26.5 & 24.0 \\
             \bottomrule
        \end{tabular}
        \caption{{Daily granularity example extracted from the Wikipedia workload. The numbers are in million.}}
        \label{tab:daily_dataset}
    \end{subtable}
    
    \begin{subtable}{0.48\textwidth}
        \centering
        \begin{tabular}{ccccccccc}
        \toprule
             \textbf{Workload} & \textbf{Week} & \textbf{M} & \textbf{Tu} & \textbf{...} & \textbf{Sa} & \textbf{Su} \\
             \toprule
            Wikipedia & 2023-01-01 & 541.5 & 533.8 & ... & 527.6 & 556.1 \\
            Wikipedia & 2023-01-08 & 543.1 & 525.8 & ... & 531.4 & 581.3 \\
            Wikipedia & 2023-01-15 & 567.2 & 549.5 & ... & 518.2 & 566.9 \\
             \bottomrule
        \end{tabular}
        \caption{{Weekly granularity example extracted from the Wikipedia workload. The numbers are in million.}}
        \label{tab:weekly_dataset}
    \end{subtable}
    
    \label{tab:daily_weekly_datasets}
\end{table}

\noindent \textbf{Daily Granularity:}
\label{sec:preprocessing_daily}
Analyzing daily granularity allows for a micro-level understanding of user behavior and system performance. This examination offers detailed insights into user engagement, system load, and operational peaks over a 24-hour cycle. This granularity enables the detection of short-term trends, such as hourly spikes in traffic or variations in user engagement between weekdays and weekends.

\begin{figure}[t]
  \centering
  \begin{subfigure}{0.24\textwidth}
    \includegraphics[width=\textwidth]{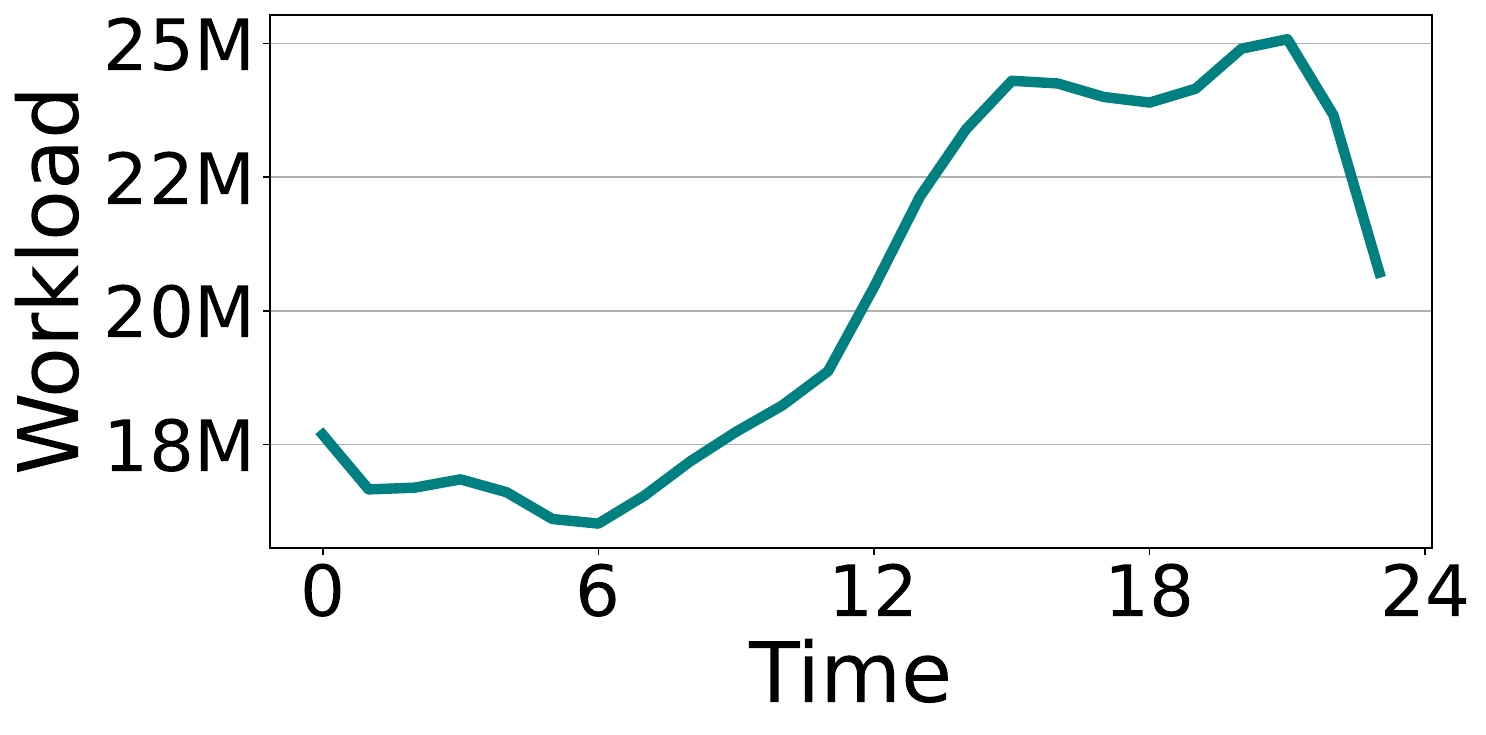}
    \caption{Wikipedia}
    \label{fig:wikipedia_hour}
  \end{subfigure}
  \hfill
  \begin{subfigure}{0.24\textwidth}
    \includegraphics[width=\textwidth]{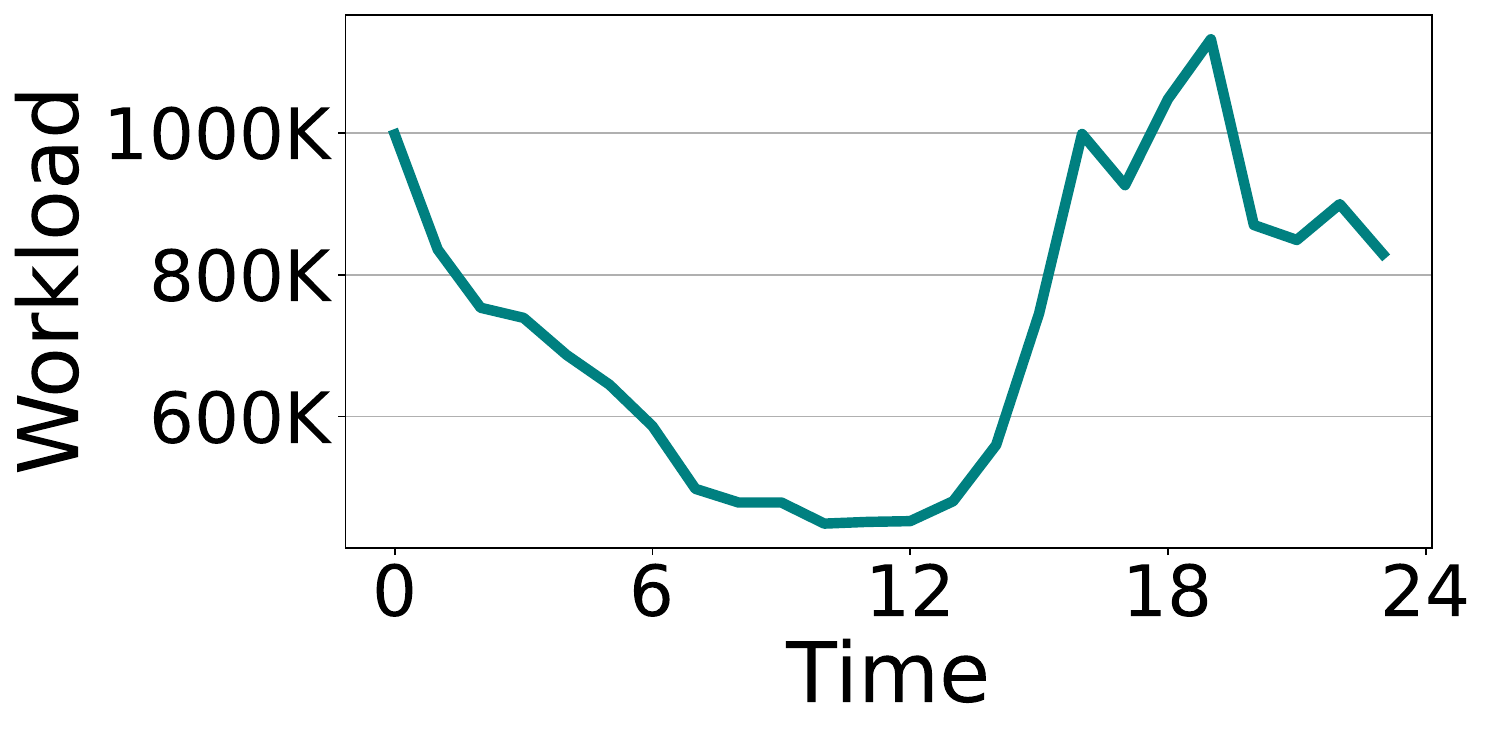}
    \caption{Worldcup98}
    \label{fig:wc98_hour}
  \end{subfigure}
  \caption{Daily granularity of Wikipedia and Worldcup98 workloads over a one-day time span}
  \label{fig:workloads_hour}
\end{figure}

We aggregate workloads in one-hour time intervals. Thus, if a single workload spans one day, it will be represented in 24 instances, each corresponding to the number of user accesses in a 1-hour window. We provide an example of the daily granularity in Table~\ref{tab:daily_dataset}, where each row corresponds to one day of data. Figure~\ref{fig:workloads_hour} presents the Wikipedia and Worldcup98 workloads in daily granularity, where each data point signifies the workload for one hour. As evident, patterns observed within these two days exhibit both similarities and differences, and the ultimate objective of this work is to uncover and understand these patterns within web application workloads.

\noindent \textbf{Weekly Granularity:}
\label{sec:preprocessing_weekly}
Weekly granularity offers a broader perspective, capturing trends and variations that span across different days of the week. Analyzing these patterns allows for the identification of cyclical behaviors and longer-term trends. For instance, it reveals differences in user engagement and system load along different seasons of the year.

\begin{figure}[t]
  \centering
  \begin{subfigure}{0.24\textwidth}
    \includegraphics[width=\textwidth]{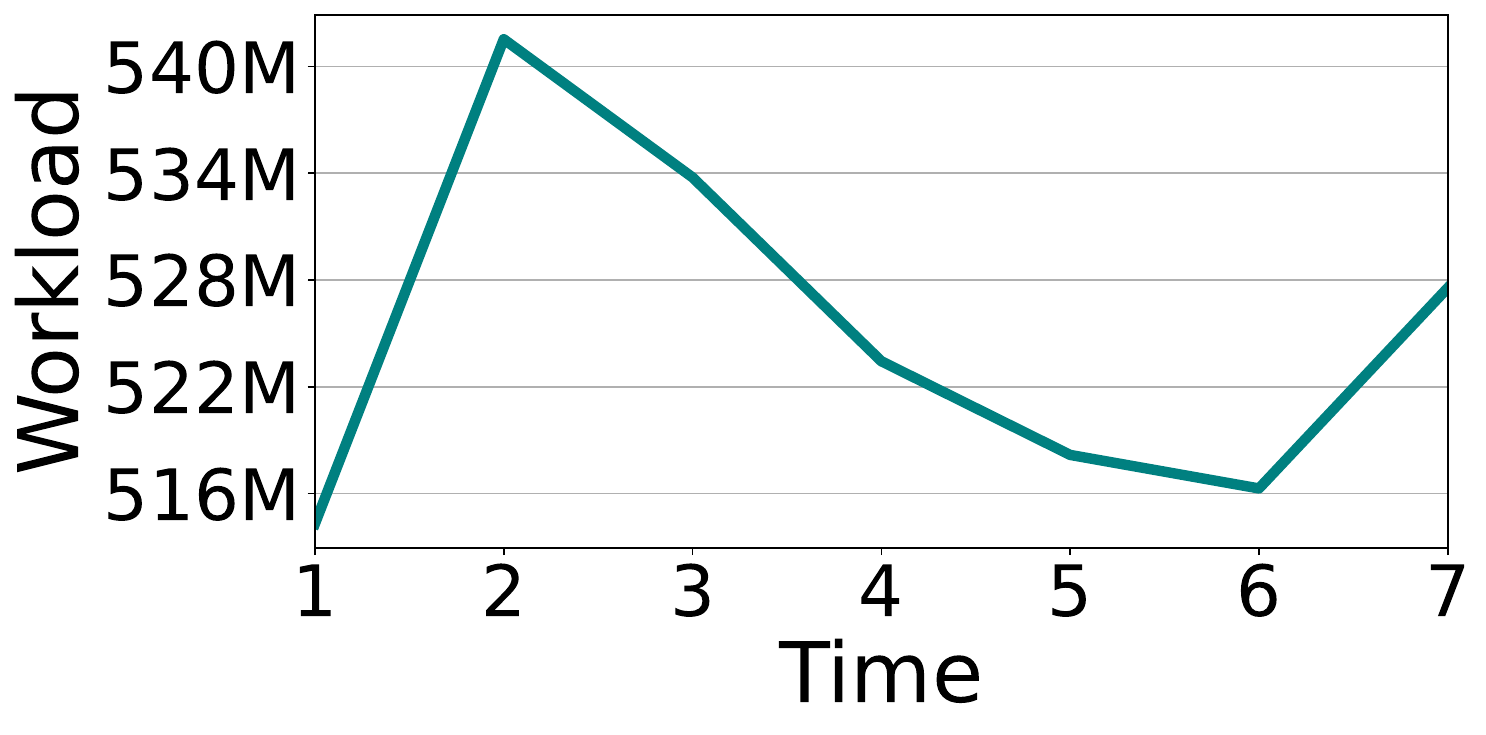}
    \caption{Wikipedia}
    \label{fig:wikipedia_day}
  \end{subfigure}
  \hfill
  \begin{subfigure}{0.24\textwidth}
    \includegraphics[width=\textwidth]{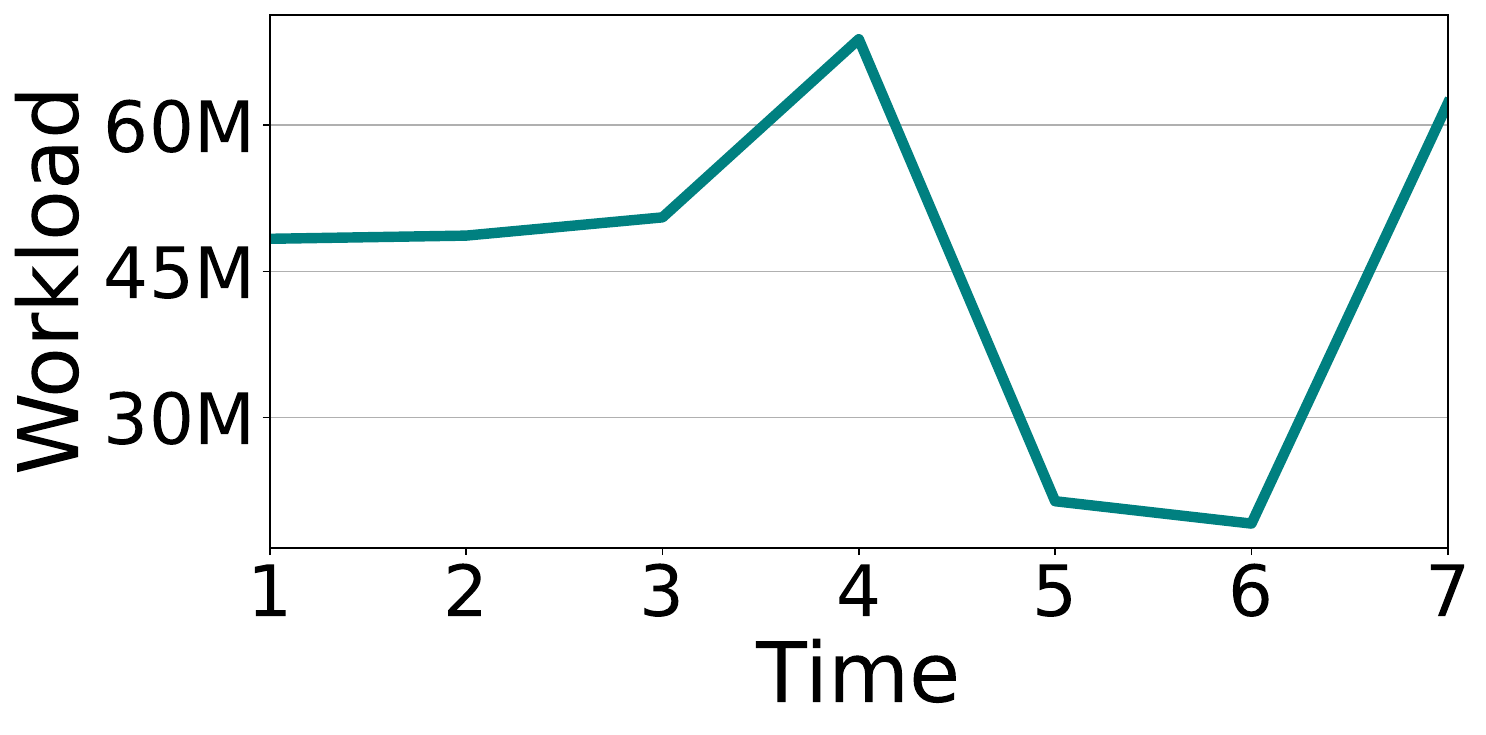}
    \caption{Worldcup98}
    \label{fig:wc98_day}
  \end{subfigure}
  \caption{Weekly granularity of Wikipedia and Worldcup98 workloads over a one-week time span}
  \label{fig:workloads_day}
\end{figure}

To analyze weekly granularity, we aggregate the workloads in time intervals of one day. In this case, if a workload encompasses a duration of one week, it will be reflected in seven instances, each representing the number of user accesses in a 1-day window. We provide an example of the weekly granularity in Table~\ref{tab:weekly_dataset}, where each row corresponds to one week of data. Figure~\ref{fig:workloads_day} shows the Wikipedia and Worldcup98 workloads in weekly granularity, where each data point signifies the workload for one day.

\paragraph{Standardization}
We standardize our data to ensure uniformity across different workloads, especially when they exhibit significant load variations. For example, in Figure~\ref{fig:workloads_day}, we observe a considerable difference in scale between Wikipedia and Worldcup98 workloads: Wikipedia’s workload scale is more than 10 times greater compared to that of Worldcup98. Thus, standardization is essential to ensure that the subsequent clustering is not biased by different workload intensities.

Z-score is a widely used technique that adjusts data to have a mean of 0 and a standard deviation of 1. The formula applies to each data point in both daily and weekly granularities.

\begin{equation}
z = \frac{X(t) - \mu}{\sigma}
\end{equation}
where \( \text{X}(t) \) represents the original data value at time \( t \), and $\mu$ and $\sigma$ represent the mean and standard deviation, respectively.

\paragraph{Smoothing}
Smoothing is employed to reduce noise and fluctuations in the workload data, thereby enhancing its clarity and facilitating trend identification. One commonly used smoothing technique is \textit{Exponential Moving Average (EMA)}. The main benefits of using the exponential smoothing method are its low cost and ease of application~\cite{karmaker2017determination}. It assigns exponentially decreasing weights to past observations, ensuring that recent data points have a higher influence on the smoothed value compared to older ones. The EMA smoothing formula is as follows:
\begin{equation}
\text{EMA}(t) = \alpha \times \text{X}(t) + (1 - \alpha) \times \text{EMA}(t - 1)
\end{equation}
where \( \text{EMA}(t) \) represents the smoothed value at time \( t \), \( \text{X}(t) \) denotes the original data value at time \( t \), and 
\( \alpha \) is the smoothing factor, typically a value between 0 and 1, determining the weight assigned to the current observation.

\subsubsection{Variability and burstiness analysis}
Variability and burstiness analysis is performed before clustering to understand the inherent dynamics of web application workloads. This analysis helps identify distribution differences that might influence clustering results. By examining raw datasets for variability and burstiness after aggregation but before standardization, we retain the original data characteristics.

Variability is defined as the extent to which data points in a time series differ from each other and from the mean value~\cite{wu2007trend}. We calculate variability by measuring the \textit{coefficient of variation (CV)} of the workload intensities within one day (i.e., 24 hourly workload intensity values) or one week (i.e., seven daily intensity values) using the below formula:

\begin{equation}
 \text{CV} = \frac{\sigma}{\mu}
\end{equation}

where \( \sigma \) is the standard deviation and \( \mu \) is the mean. 

Burstiness refers to the tendency of a time series to exhibit sudden, irregular increases in activity or intensity over short periods. To quantify burstiness, we calculate the mean and standard deviation of workloads per day and per week. Burstiness is then determined using the formula~\cite{goh2008burstiness}:

\begin{equation}
\text{Burstiness} = \frac{\sigma - \mu}{\sigma + \mu} 
\end{equation}

\begin{figure}[t]
  \centering
  \begin{subfigure}{0.24\textwidth}
    \includegraphics[width=\textwidth]{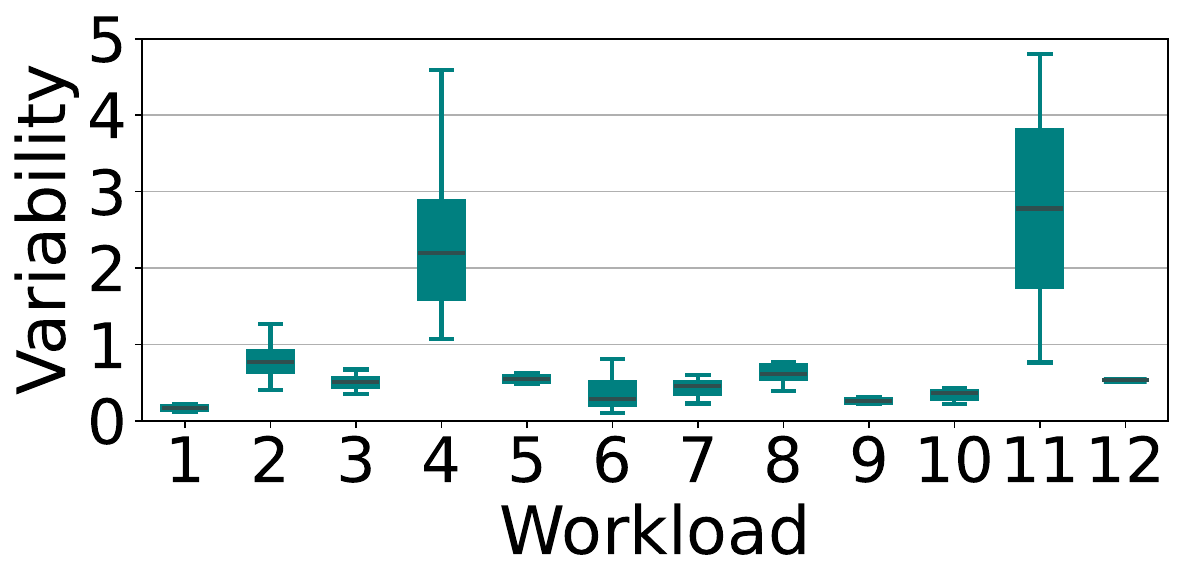}
    \caption{Daily granularity}
    \label{fig:variance_daily}
  \end{subfigure}
  \hfill
  \begin{subfigure}{0.24\textwidth}
    \includegraphics[width=\textwidth]{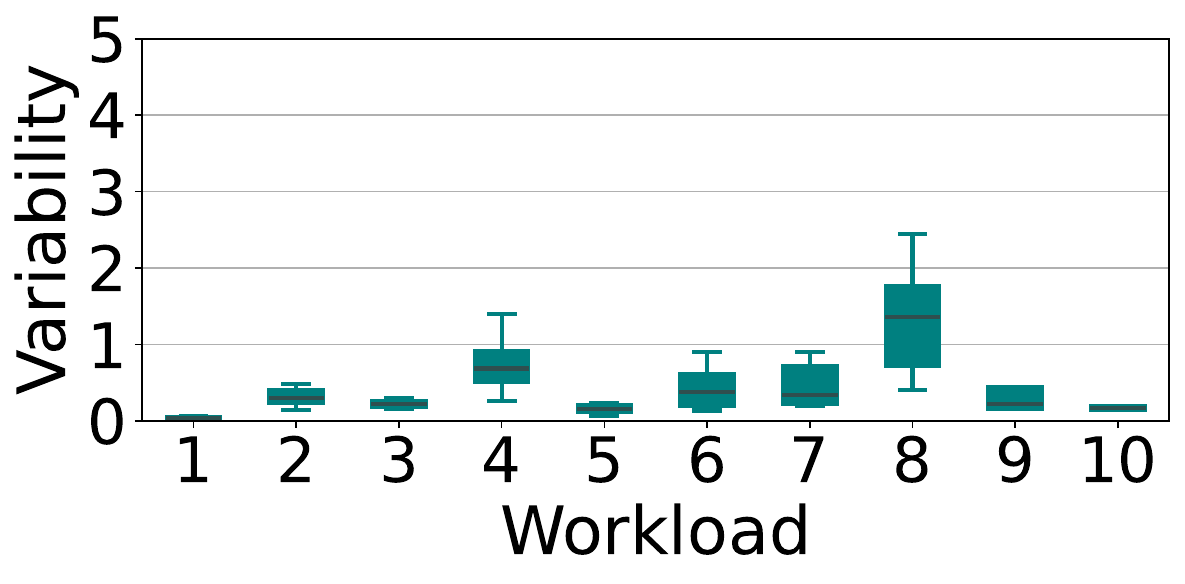}
    \caption{Weekly granularity}
    \label{fig:variance_weekly}
  \end{subfigure}
  \caption{Variability analysis of daily and weekly granularities}
  \label{fig:variability}
\end{figure}

\begin{figure}[t]
  \centering
  \begin{subfigure}{0.24\textwidth}
    \includegraphics[width=\textwidth]{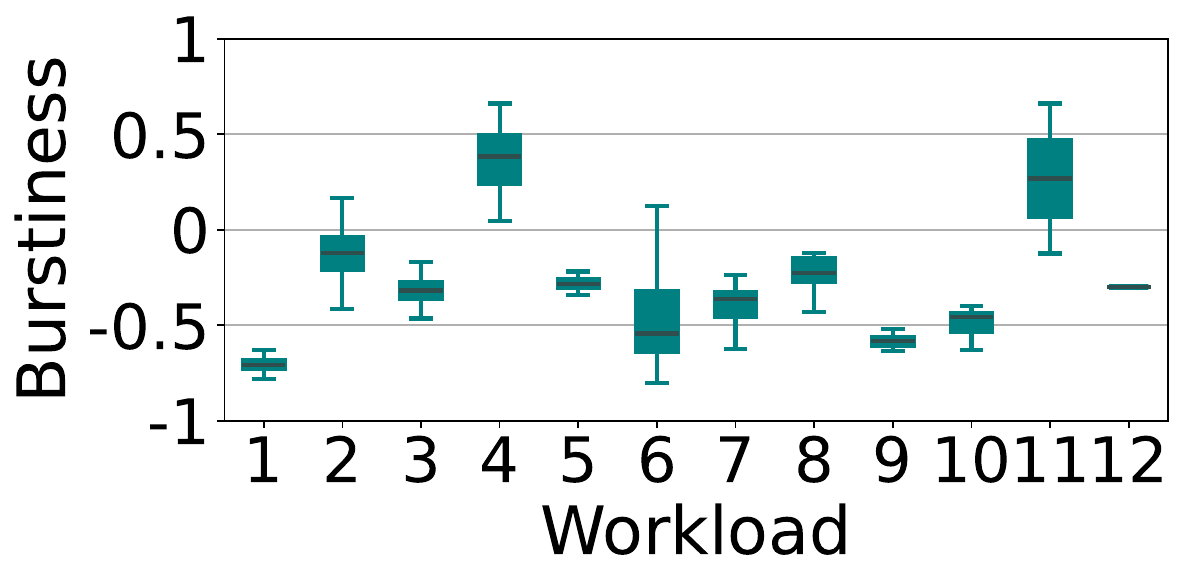}
    \caption{Daily granularity}
    \label{fig:burstiness_daily}
  \end{subfigure}
  \hfill
  \begin{subfigure}{0.24\textwidth}
    \includegraphics[width=\textwidth]{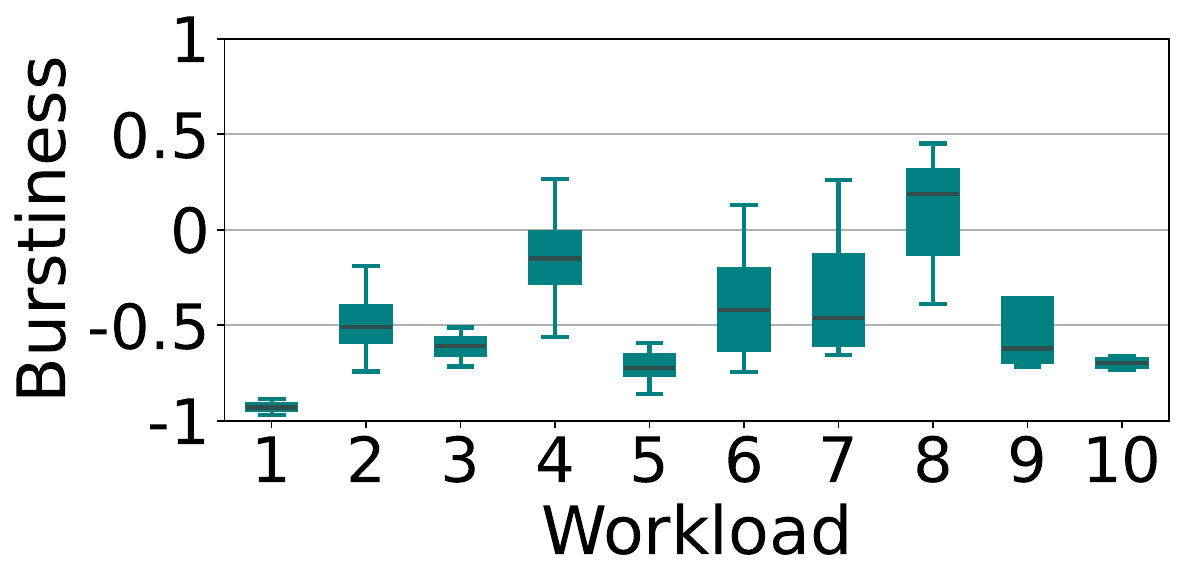}
    \caption{Weekly granularity}
    \label{fig:burstiness_weekly}
  \end{subfigure}
  \caption{Burstiness analysis of daily and weekly granularities}
  \label{fig:burstiness}
\end{figure}

Burstiness ranges from -1 to 1, where 1 indicates high irregularity, 0 indicates no significant burstiness, and -1 suggests a regular pattern. This quantification highlights periods with significant deviations from overall trends.

\subsubsection{Clustering the Workloads}
\label{sec:clustering}
To investigate general workload patterns across a comprehensive dataset, we employ workload clustering. We first develop a script to combine the standardized and smoothed daily and weekly workloads from all 12 workload datasets. This combination process results in two unified datasets: the daily workload dataset and the weekly workload dataset. This combination enables us to construct a unified dataset that can effectively capture robust and generalized patterns. The daily dataset contains 3191 data points, each representing one day, and the weekly dataset contains 466 data points, each representing one week. 

After constructing the combined daily and weekly datasets, we employ K-Means clustering. K-Means clustering is a widely recognized and employed clustering method, designed to divide \textit{n} observations into \textit{k} clusters, in which analyzed data sets are partitioned in relation to the selected parameters and grouped around cluster centroids~\cite{xu2005survey}. 

Determining the optimal number of clusters (a.k.a., \textit{k} value) is critical in applying the K-Means algorithm. Various methods exist for this purpose, including the elbow method~\cite{syakur2018integration} and the silhouette score~\cite{shahapure2020cluster}. In this study, we opt for the silhouette score, which measures the cohesion and separation of clusters, with values ranging from -1 to 1. A higher value indicates well-separated clusters and a near-zero or negative score suggests overlapping or incorrectly assigned data points.

We experiment with \textit{Euclidean}~\cite{danielsson1980euclidean}, \textit{Dynamic Time Warping (DTW)}~\cite{muller2007dynamic}, and \textit{Soft-DTW}~\cite{cuturi2017soft} distance metrics to determine the most suitable metric for our data. Our evaluation indicates that Euclidean distance yieldes the most meaningful cluster separations based on the silhouette score. We execute the K-Means algorithm for \textit{k} values ranging from 1 to 20 and determine the optimal \textit{k} based on the silhouette score. Additionally, to validate our findings, we visually inspect the clustering results using \textit{t-SNE} illustration~\cite{van2008visualizing}.

\subsubsection{Analyzing Cluster Characteristics}
Beyond identifying workload patterns, understanding their underlying characteristics and interrelationships is crucial. We analyze the clustering results from various perspectives: analyzing centroids, studying associations between daily and weekly patterns, and exploring the time dependence of workload patterns.

\paragraph{Centroid Analysis}
A centroid is a representative point that summarizes the central tendency of the data and provides insight into its distribution or trend over time. In our analysis, we compute centroid values for each daily and weekly pattern, aiming to capture the characteristics of each cluster's temporal behavior. To effectively model these centroids, we use \textit{Polynomial Models}, specifically \textit{Quadratic} and \textit{Cubic} polynomials, chosen for their ability to capture the nuanced shapes and fluctuations observed in workload centroids. The quadratic model is mathematically expressed as:

\begin{equation}
at^2 + bt + c = 0
\end{equation}
whereas the cubic model is expressed as:
 
 \begin{equation}
at^3 + bt^2 + ct + d = 0
\end{equation}
where \( a \), \( b \), \( c \), and \( d \) represent the coefficients of the models, and \( t \) denotes the time variable.

When fitting the polynomial models, we optimize the algorithm using the Levenberg-Marquardt algorithm~\cite{ranganathan2004levenberg}. This method continuously optimizes the model parameters to reduce the total squared differences between the model's predictions and the real data points. By doing so, this objective function guarantees that the adapted models precisely choose the temporal dynamics observed within the clusters.

\paragraph{Association between daily and weekly patterns}
We determine the associations between daily and weekly patterns by calculating the presence of daily patterns within each weekly pattern. Specifically, for each weekly pattern, we calculate the percentage of the instances associated with each daily pattern.
Through quantifying the frequency of daily patterns within weekly patterns, our objective is to unveil the hierarchical arrangement of temporal dynamics. This analytical strategy facilitates the interpretation of temporal associations, hence providing detailed insights into the patterns.

\paragraph{Time dependence of workload patterns}
We explore the association between workload patterns and time. For daily patterns, we consider their association with days of the week (i.e., weekdays or weekends), and for weekly patterns, we consider their association with seasons of the year. Specifically, we calculate the percentage of each workload pattern's presence within weekdays/weekends or each season.

\subsection{Results}
We first present the variability and burstiness analysis. Then, we provide the clustering findings in four parts: clustering patterns, centroid analysis, association between daily and weekly patterns, and time dependence of workload patterns.

\subsubsection{Variability and burstiness analysis}
Figure~\ref{fig:variability} illustrates the variability analysis for daily and weekly granularities across each workload. The number of workloads depicted in this figure (from 1 to 12) corresponds to the ``ID'' column listed in Table~\ref{tab:workloads}. A CV close to 0 indicates low variability, suggesting stable workloads, while a CV greater than 1 suggests high relative variability.  Figure~\ref{fig:variability} reveals two key observations: 1. Daily variability is significantly higher than weekly variability, and 2. Although most workloads exhibit relatively stable workloads, high variability is observed in two daily workloads (i.e., Boston and EPA) and one weekly workload (i.e., YouTube).

\begin{figure}[t]
\centering
  \begin{subfigure}{0.48\linewidth}
    \centering
    \includegraphics[width=1\linewidth]{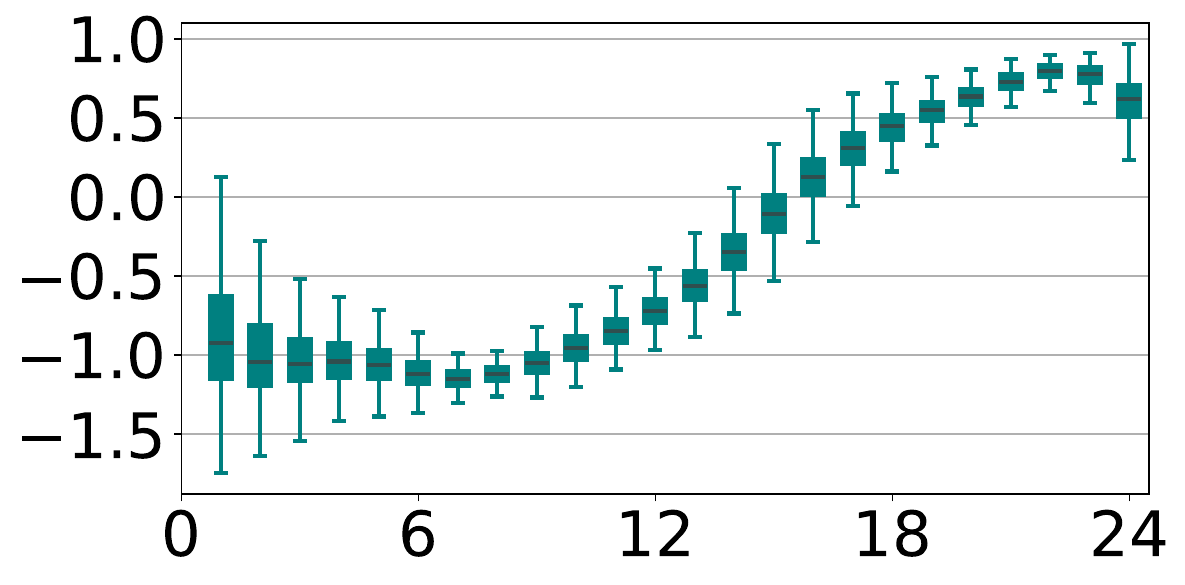}
    \caption{D1}
    \label{fig:cluster_d1}
  \end{subfigure}
    \begin{subfigure}{0.48\linewidth}
    \centering
    \includegraphics[width=1\linewidth]{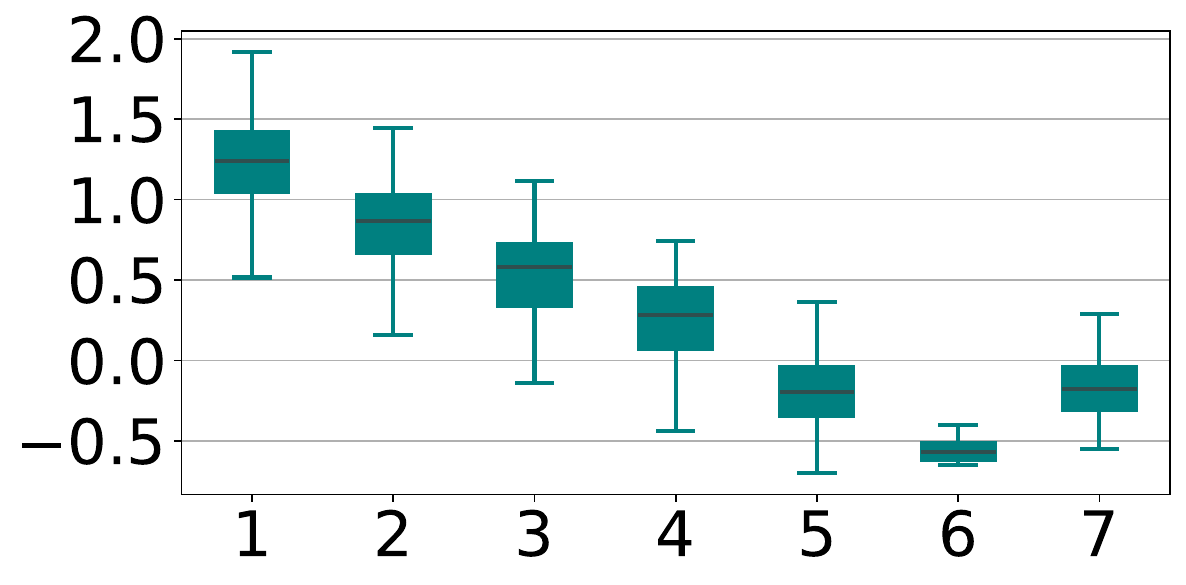}
    \caption{W1}
    \label{fig:cluster_w1}
  \end{subfigure}
    \begin{subfigure}{0.48\linewidth}
    \centering
    \includegraphics[width=1\linewidth]{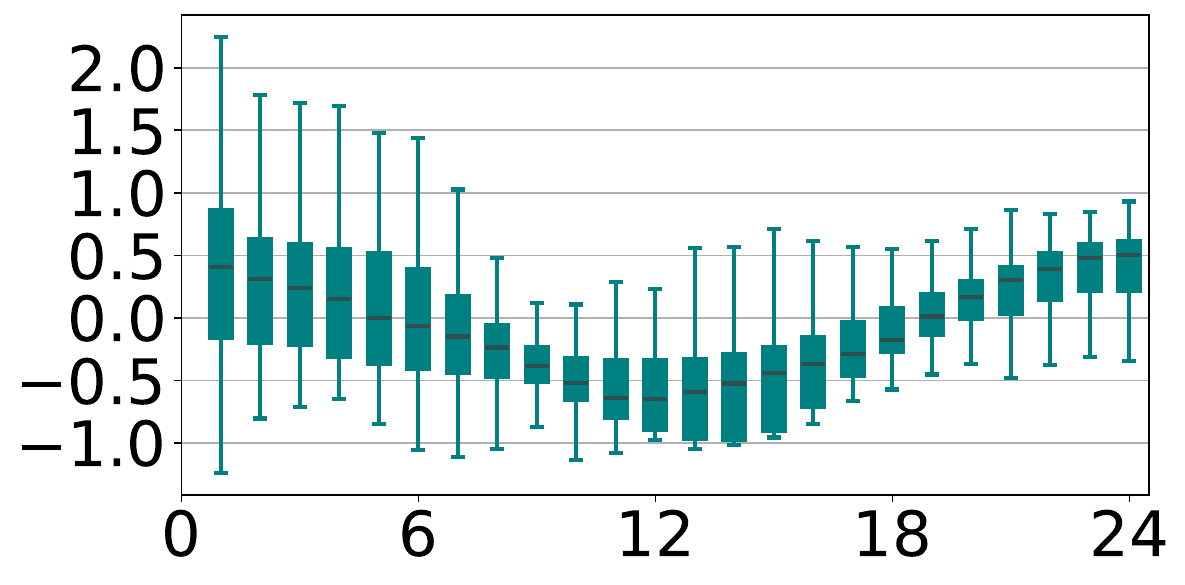}
    \caption{D2}
    \label{fig:cluster_d2}
  \end{subfigure}
    \begin{subfigure}{0.48\linewidth}
    \centering
    \includegraphics[width=1\linewidth]{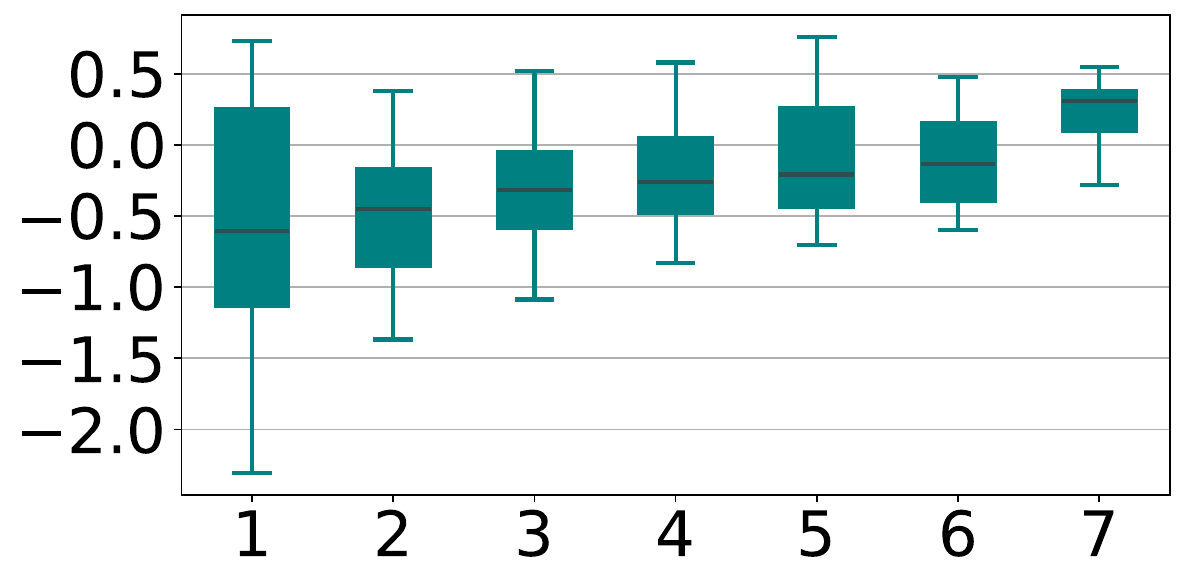}
    \caption{W2}
    \label{fig:cluster_w2}
  \end{subfigure}
    \begin{subfigure}{0.48\linewidth}
    \centering
    \includegraphics[width=1\linewidth]{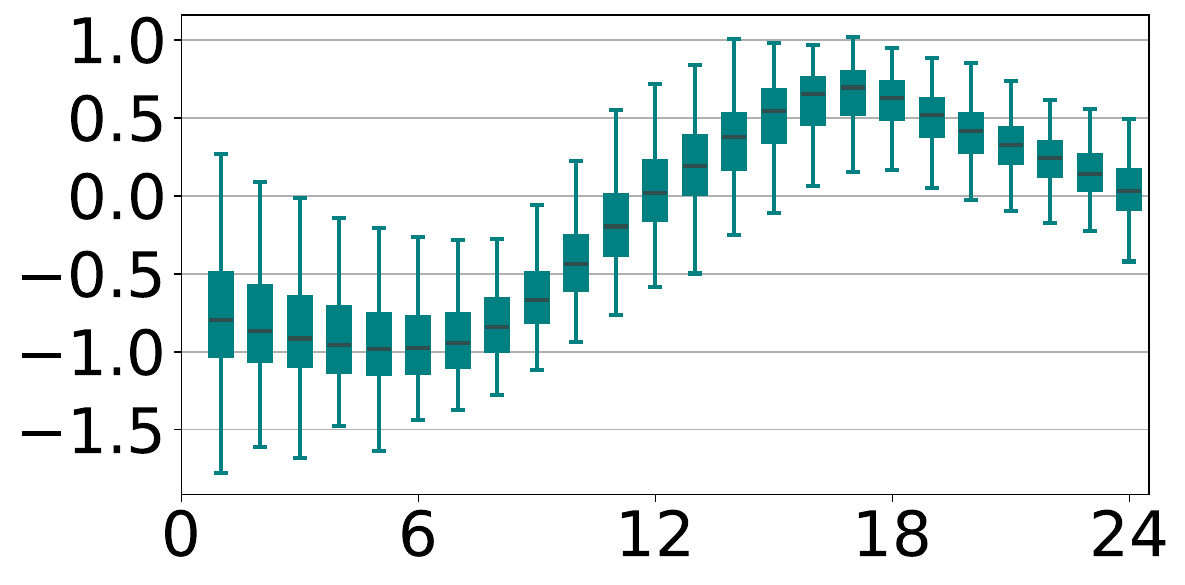}
    \caption{D3}
    \label{fig:cluster_d3}
  \end{subfigure}
  \begin{subfigure}{0.48\linewidth}
    \centering
    \includegraphics[width=1\linewidth]{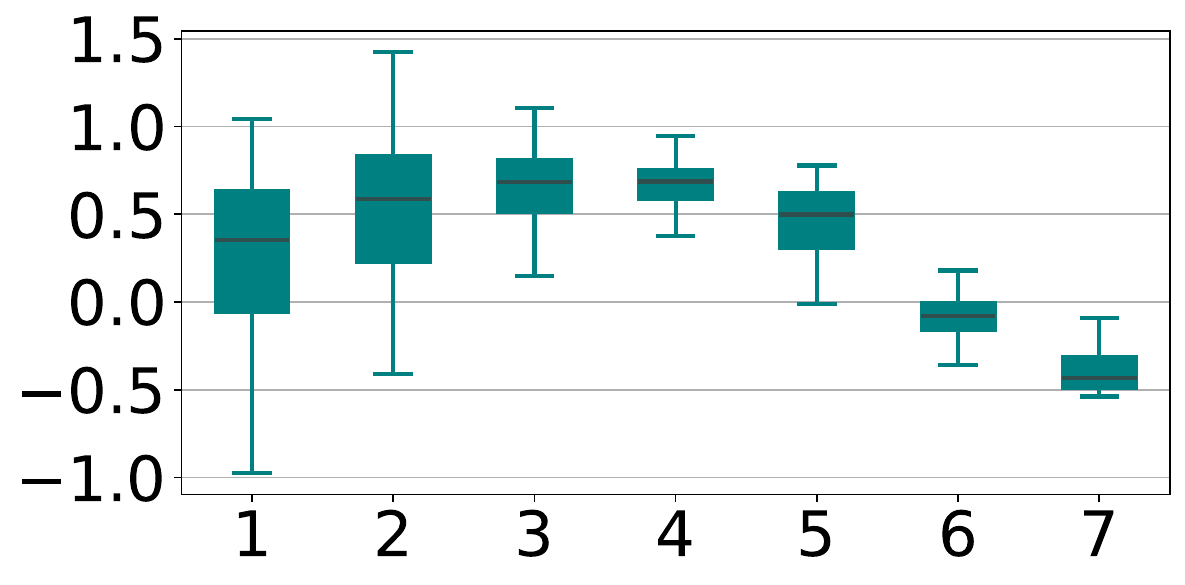}
    \caption{W3}
    \label{fig:cluster_w3}
  \end{subfigure}
  \caption{Distribution of daily and weekly clusters}
  \label{fig:daily_weekly_clusters}
\end{figure}

Figure~\ref{fig:burstiness} presents the burstiness analysis for daily and weekly granularities. Similar to variability findings, daily workloads exhibit higher burstiness. Specifically, while none of the weekly workloads exhibit burstiness values exceeding 0.5, two of the daily workloads, Boston and EPA, reach this threshold. The Wikipedia workload shows the lowest burstiness, nearing -1, indicating regular patterns. This regularity may result from its large global user base, which balances bursts across different regions.

\begin{mybox}
There exist significant variations and burstiness within the workloads of studied web applications, which should be taken into account by practical resource provisioning and workload generation strategies.
\end{mybox}

\subsubsection{Clustering Patterns}
The clustering results are presented in Figure~\ref{fig:daily_weekly_clusters}, showcasing the identified patterns. As mentioned in Section~\ref{sec:clustering}, we obtain 3191 datapoints at the daily granularity of web application workloads. After clustering, we identify three distinct patterns. Specifically, Figure~\ref{fig:cluster_d1} illustrates the first cluster (i.e., D1) with 2262 instances, Figure~\ref{fig:cluster_d2} shows the second cluster (i.e., D2) with 406 instances, and Figure~\ref{fig:cluster_d3} depicts the third cluster (i.e., D3) with 523 instances. The x-axis of these figures represents a 24-hour day.

As discussed in Section~\ref{sec:clustering}, 466 datapoints exist at the weekly granularity of web application workloads. After clustering, we uncover three patterns presented in Figures~\ref{fig:cluster_w1}, ~\ref{fig:cluster_w2}, and~\ref{fig:cluster_w3}, and are donated as W1, W2, and W3, with 283, 64, and 119 instances, respectively. The x-axis of these figures corresponds to a 7-day week, commencing from 1 (Monday) and concluding at 7 (Sunday).

Looking at daily and weekly patterns, it is evident that they exhibit unique patterns while having commonalities. Most of the clusters (i.e., D1, D2, D3, W1, W3) are non-monotonic and have curvy shapes. On the other hand, W2 has a sublinear shape with a slightly ascending pattern. All the patterns show a distinct peak and low period, yet the rate of ascending/descending to/from these peaks differs. 


\begin{mybox}
Our clustering reveals three distinct patterns for both daily and weekly workloads. Interestingly, most of these patterns are non-monotonic and non-linear. 
\end{mybox}



\subsubsection{Centroid Analysis}
\begin{figure}[t]
\centering
  \begin{subfigure}{0.48\linewidth}
    \centering
    \includegraphics[width=1\linewidth]{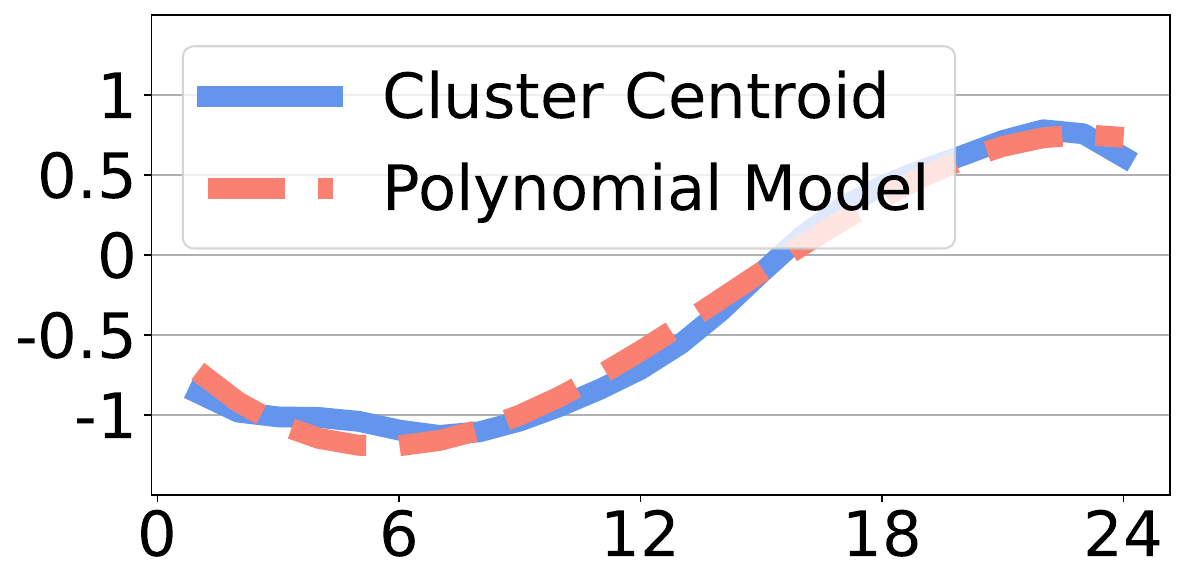}
    \caption{D1, a = -0.001, b = 0.029, \\ c = -0.221, d = -0.728}
    \label{fig:centroid_cluster_d1}
  \end{subfigure}
    \begin{subfigure}{0.48\linewidth}
    \centering
    \includegraphics[width=1\linewidth]{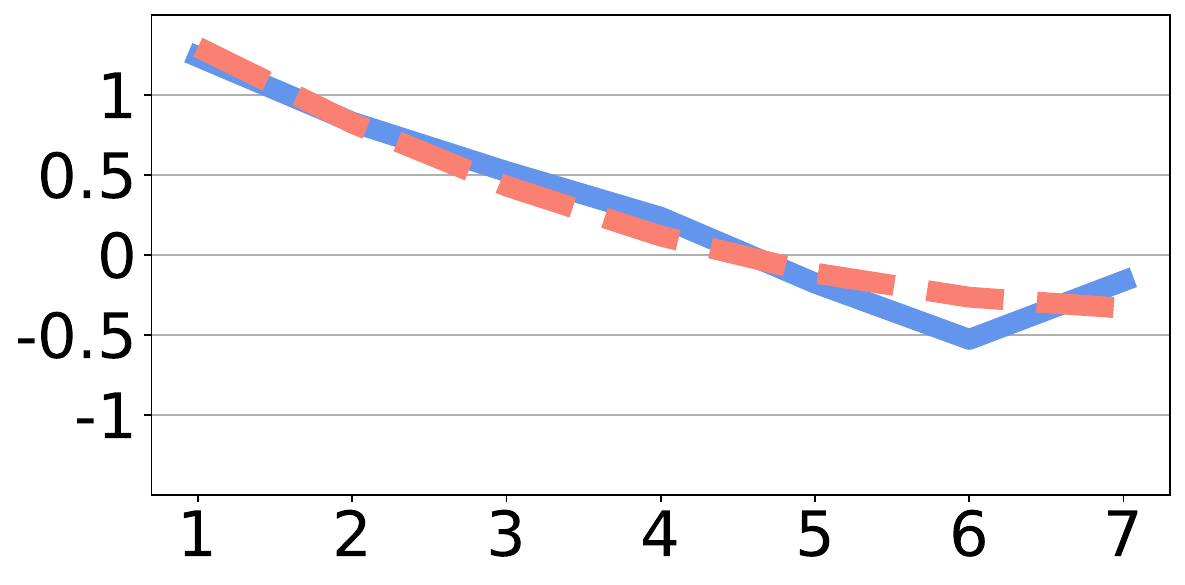}
    \caption{W1, a = 0.041, b = -0.516, \\ c = 1.299}
    \label{fig:centroid_cluster_w1}
  \end{subfigure}
    \begin{subfigure}{0.48\linewidth}
    \centering
    \includegraphics[width=1\linewidth]{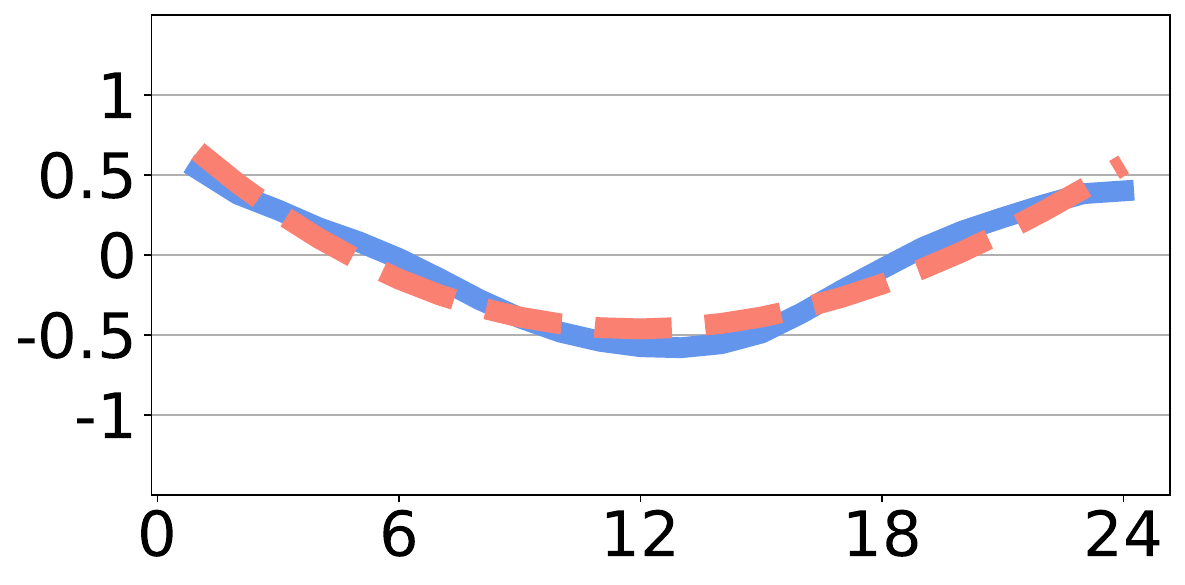}
    \caption{D2, a = 0.000, b = 0.011, \\ c = -0.214, d = 0.648}
    \label{fig:centroid_cluster_d2}
  \end{subfigure}
    \begin{subfigure}{0.48\linewidth}
    \centering
    \includegraphics[width=1\linewidth]{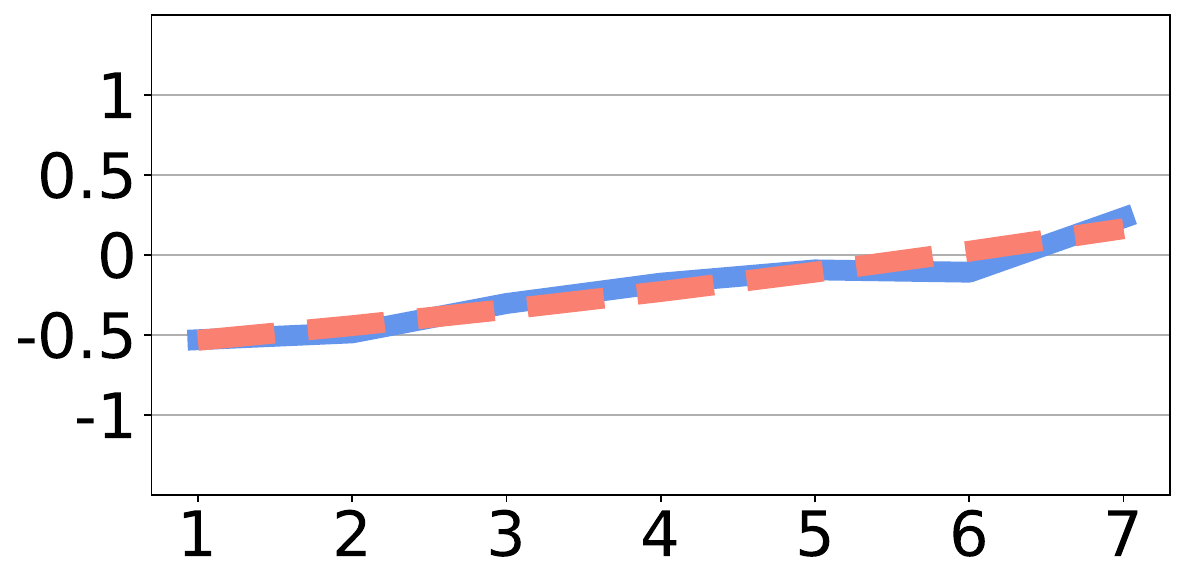}
    \caption{W2, a = 0.005, b = 0.087, \\ c = -0.535}
    \label{fig:centroid_cluster_w2}
  \end{subfigure}
    \begin{subfigure}{0.48\linewidth}
    \centering
    \includegraphics[width=1\linewidth]{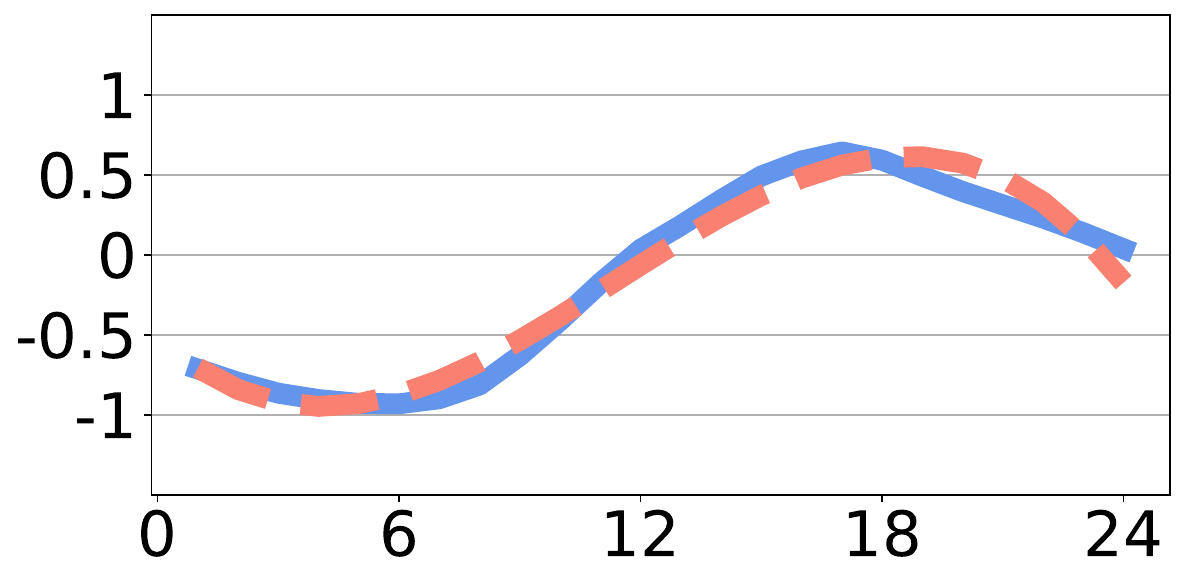}
    \caption{D3, a = -0.001, b = 0.031, \\ c = -0.166, d = -0.708}
    \label{fig:centroid_cluster_d3}
  \end{subfigure}
  \begin{subfigure}{0.48\linewidth}
    \centering
    \includegraphics[width=1\linewidth]{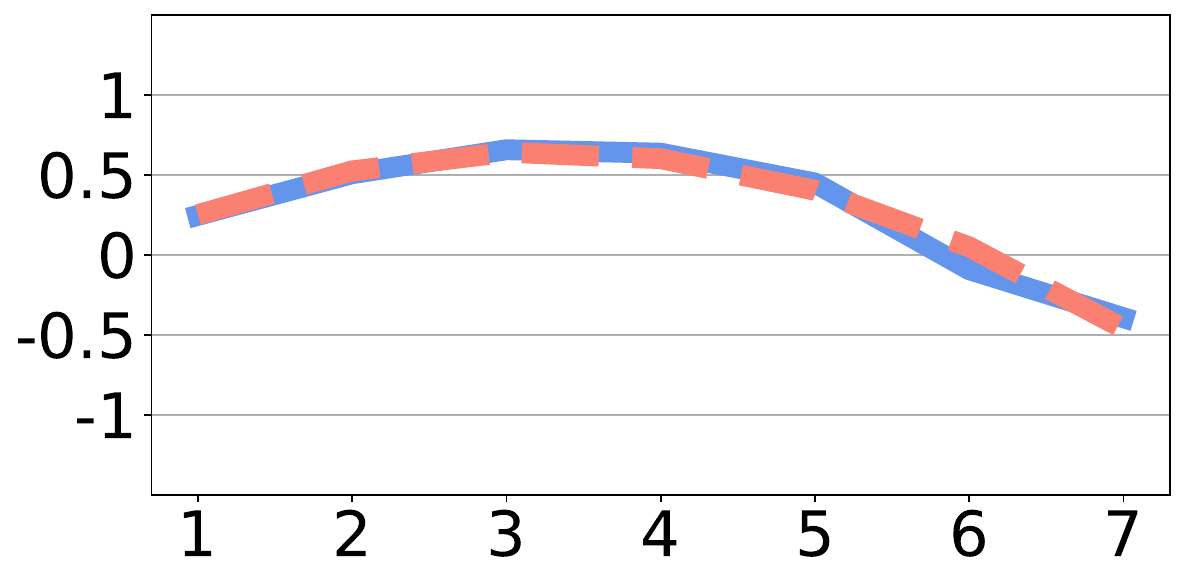}
    \caption{W3, a = -0.079, b = 0.352, \\ c = 0.251}
    \label{fig:centroid_cluster_w3}
  \end{subfigure}
  \caption{Daily and weekly centroids}
  \label{fig:daily_weekly_centroids}
\end{figure}

Figure~\ref{fig:daily_weekly_centroids} shows the centroids of daily and weekly patterns. These centroids serve as fundamental representations of the underlying temporal dynamics captured within each cluster. Utilizing both quadratic polynomial and cubic polynomial approaches, we discover that the cubic polynomial model best fits our daily clusters and can successfully model the underlying nuances. On the other hand, for weekly patterns, due to their simpler shapes, the quadratic model is enough to fit the centroids. The coefficients of the cubic model (for daily patterns) and the quadratic model (for weekly patterns) are presented in~Figure~\ref{fig:daily_weekly_centroids}.

Complementing our mathematical modeling, we assign descriptive names to the clusters based on their centroid patterns.

\textbf{Daytime active with rapid decline (D1)}: Substantial activity during daytime hours with a rapid decrease in activity from peak to off-peak hours.

\textbf{Nighttime active (D2)}: Primary activity during the nighttime with gradual workload changes.

\textbf{Daytime active with gradual decline (D3)}: Substantial activity during daytime hours with a gradual decrease in activity from peak to off-peak hours.

\textbf{Weekday Active (W1)}: Higher activity on weekdays with consistent decrease.

\textbf{Weekend Active (W2)}: Increased activity on weekends with steady rise.

\textbf{Midweek Active (W3)}: Activity concentrated in midweek, exhibiting smooth changes.

\begin{mybox} The daily and weekly patterns follow polynomial (i.e., cubic and quadratic) models. These patterns represent varying levels of user activity related to daytime/nighttime and weekdays/weekends.
\end{mybox}

\subsubsection{Association between daily and weekly patterns}
Table~\ref{tab:frequency} shows the relative frequency of daily and weekly patterns. Notably, D1 emerges as the predominant daily pattern, while W1 stands out as the most common weekly pattern. Moreover, the combination of D1 and W1 represents the most frequent occurrence, encompassing over 43\% of the workloads. In Table~\ref{tab:frequency}, we include the count of workload datasets associated with each pattern within parentheses. For instance, the combination of D1 and W1 appears in six workload datasets. This observation suggests that the identified patterns are not specific to individual datasets, rather, they exist in multiple workload datasets. 

\begin{mybox}
Each weekly pattern has a dominant (i.e., $>$50\%) daily pattern; D1 for W1 and W2, D3 for W3. Besides, the identified patterns are general across different workload datasets.
\end{mybox}

\begin{table}[t]
\caption{Relative frequency of daily and weekly clusters, expressed in percentage. The numbers in parentheses indicate the frequency of these patterns across workload datasets.}
\centering
\footnotesize
    \begin{tabular}{cccccc}
    \toprule
         \textbf{} & \textbf{D1} & \textbf{D2} & \textbf{D3} & \textbf{Total} \\
         \toprule
        \textbf{W1} & 43.6 (6) & 2.4 (5) & 3.8 (7) & 49.8 (9) \\
        \textbf{W2} & 7.3 (8) & 2.7 (9) & 3.2 (6) & 13.2 (10) \\
        \textbf{W3} & 7.8 (7) & 9.4 (10) & 19.8 (9) & 37 (12) \\
        \midrule
        \textbf{Total} & 58.7 (9) & 14.5 (10) & 26.8 (11) & 100 (12) \\
        \bottomrule
    \end{tabular}
    \label{tab:frequency}
\end{table}

\subsubsection{Time dependence of workload patterns}
\begin{figure}[t]
\centering
  \begin{subfigure}{0.49\linewidth}
    \centering
    \includegraphics[width=1\linewidth]{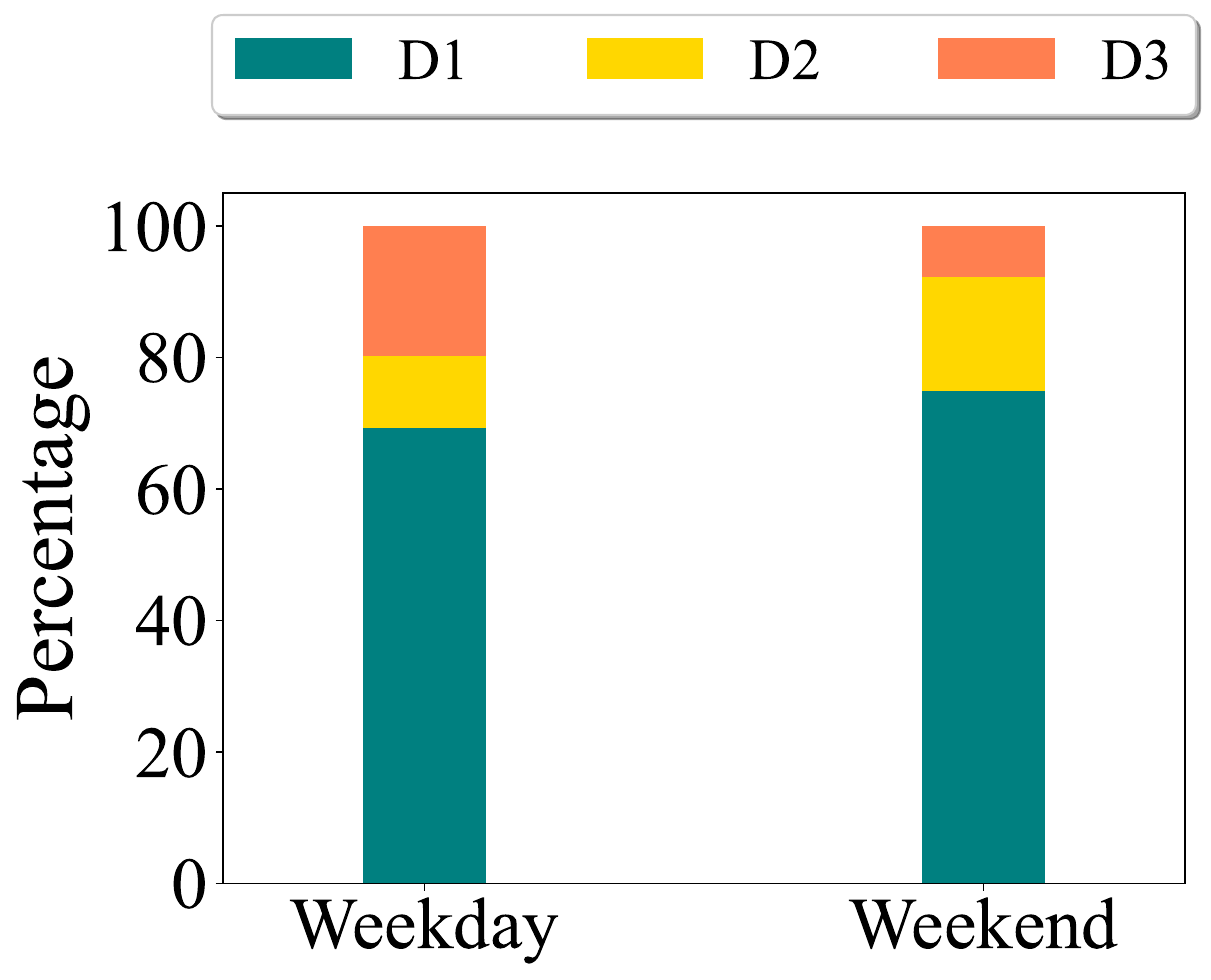}
    \caption{Daily granularity}
    \label{fig:daily_week}
  \end{subfigure}
    \begin{subfigure}{0.49\linewidth}
    \centering
    \includegraphics[width=1\linewidth]{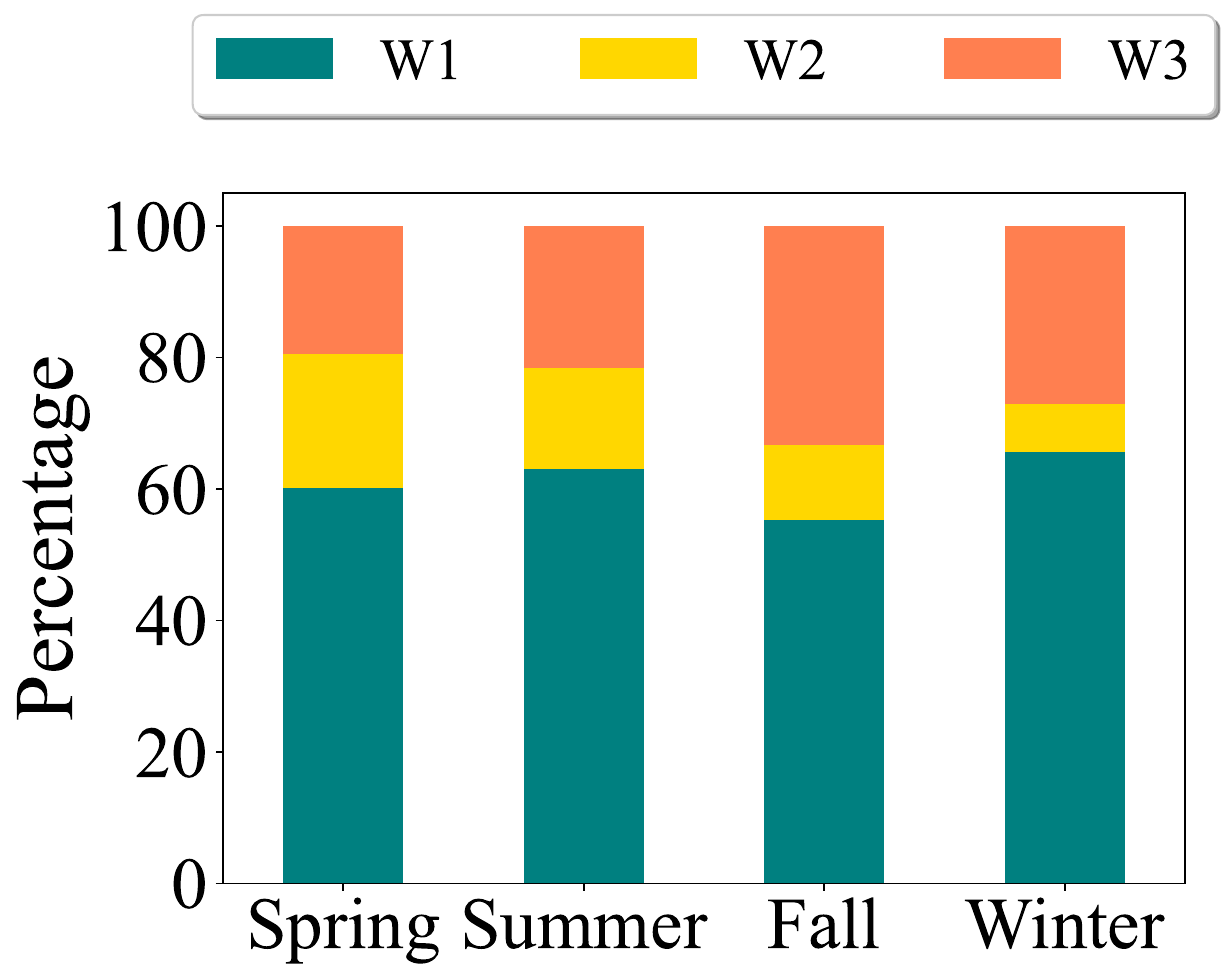}
    \caption{Weekly granularity}
    \label{fig:weekly_quarter}
  \end{subfigure}
  \caption{Composition of daily and weekly workload patterns regarding time}
  \label{fig:time}
\end{figure}

We evaluate how time affects workload patterns. Figure~\ref{fig:time} illustrates each pattern's distribution across different time periods. Figure~\ref{fig:daily_week} compares patterns between weekdays and weekends for each daily cluster. Analyzing the figure reveals that while D1 follows a typical distribution, D2 and D3 diverge from this trend, with D2 showing more weekend data and D3 displaying a higher proportion of weekday data, suggesting variations in workload patterns between weekdays and weekends.

In Figure~\ref{fig:weekly_quarter}, we examine the workload pattern distribution across weekly clusters throughout the seasons. Similar to daily patterns, there are noticeable variations in the distribution of weekly patterns across different seasons. Specifically, cluster W1 shows a higher proportion during the Summer and Winter periods, whereas cluster W3 displays more variability, with a notable increase in percentage during the Fall season.

\begin{mybox}
Variations in workload patterns across different time periods, such as weekdays versus weekends, underscore the importance of considering temporal dynamics for realistic workload generation approaches.
\end{mybox}

\section{Discussion}
\label{sec:discussion}

\textbf{Random, steady, or linearly increased/decreased workloads should be replaced by polynomially evolving workloads in realistic workload generation.}
For synthetic workload generation, researchers commonly use steady (i.e., constant level of activity) or linear (i.e., linear step-wise increase/decrease of the level of activity to model the light/normal/peak usage) patterns~\cite{jiang2015survey}. These patterns have been extensively adopted in existing work (e.g.,~\cite{masdari2020survey, persico2017fuzzy}). Some studies also utilize random workloads (e.g.,~\cite{kim2016empirical, liao2020using}). However, our RQ2 findings reveal that real-world web application workloads exhibit non-monotonic and non-linear (polynomial) behavior. Thus, we recommend that future studies consider polynomial workload patterns instead of random, linear, or steady ones. The polynomial models that we derived in this work can be directly used to guide the design of realistic workloads.


\textbf{Resource provisioning strategies can leverage the identified daily and weekly workload patterns to achieve simpler and more robust resource allocations.}
Existing proactive resource provisioning strategies usually rely on predictive models to make provisioning decisions (e.g.,~\cite{shahidinejad2021resource, zhou2022cushion}). 
However, such predictive models often suffer from the challenges of interpretability and short prediction windows~\cite{Menna2024TimeSeries}.
Practitioners can leverage the characterized workload patterns in our work to enhance the simplicity and robustness of their resource provisioning strategies. 
For example, they can progressively increase or decrease their provisioned resources based on a polynomial pattern (the parameters of the polynomial can be periodically learned from their workload data).

\textbf{We call for the sharing of newer web application workload datasets.}
Our findings indicate a tendency to use outdated workload datasets in the literature. This reliance may fail to reflect the current dynamics of real-world scenarios due to significant changes in web technologies and the exponential growth of data in recent years. We encourage researchers to utilize more up-to-date workload datasets such as Wikipedia, which better capture the behaviors of modern users. Additionally, we urge researchers and practitioners to extract and share more recent web application workloads publicly.

\section{Threats to Validity}
\label{sec:threats}

\textbf{External validity.} 
We conduct a thorough review of web application workloads, analyzing 78 articles that leverage them. To mitigate the risk of missing some studies, we conduct a systematic approach including keyword search, forward and backward snowballing, and manual analysis. However, there's a possibility of missing some studies, such as studies based on private datasets or non-English publications. 

\textbf{Internal validity.} 
Internal validity may be threatened by the assumption that daily and weekly patterns adequately capture the temporal dynamics of web application workloads. It is possible that different temporal sequences, beyond daily and weekly, could reveal alternative patterns that were not considered. We deliberately focus on daily and weekly patterns as they are widely recognized temporal units in the context of web applications. Our analysis remains open to future exploration of other temporal granularities.

\textbf{Construct validity.} 
We use the K-Means algorithm for clustering workload patterns, known for its simplicity and effectiveness, with widespread usage in previous studies (e.g., ~\cite{shahidinejad2021resource, chen2018does}). However, the choice of clustering algorithm can impact results. We encourage future research to explore alternative clustering algorithms. Moreover, the selection of the number of clusters may affect the quality and insights of the outcomes. To minimize bias, we employ a combination of quantitative metrics (i.e., silhouette score) and qualitative techniques (i.e., visualization) to determine the optimal number of clusters.

\section{Conclusion}
\label{sec:conclusion}
In this study, we perform a systematic literature review to understand the utilization and characteristics of web application workloads. Using a systematic approach, we identify 78 articles leveraging web application workloads and the 12 public workload datasets they utilize. 
While we observe that a wide spectrum of studies repetitively leverages these workload data for resource management, workload analysis, self-adaption, and benchmarking, we also notice a significant reliance on dated datasets.
Through the characterization of the 12 identified workload datasets at daily and weekly granularities, we uncover three daily and three weekly patterns. Using statistical modeling, we find that these patterns display polynomial (non-monotonic and non-linear) behaviors. Future work can use the insights gained from our characterization in realistic workload generation and resource provisioning strategies, ultimately leading to more efficient software maintenance practices such as performance optimization and capacity planning. Another extension of our study could be the possibility of exploring different temporal granularities beyond daily and weekly patterns (e.g., seasonal or yearly patterns).

\bibliographystyle{IEEEtran}
\bibliography{bibligraphy}

\end{document}